\documentclass[namedreferences,12pt]{mySolarPhysics}
\usepackage[optionalrh,linksfromyear,solaromanenum]{myspr-sola-addons} 
\usepackage{epsfig}            
\usepackage{graphicx}          
\usepackage{color}             
\usepackage[breaklinks]{hyperref}
\usepackage{breakurl}          
\ifx \doiurl    \undefined \def \doiurl#1{\href{http://dx.doi.org/#1}{\textsf{DOI}}}\fi
\ifx \adsurl    \undefined \def \adsurl#1{\href{http://adsabs.harvard.edu/abs/#1}{\textsf{ADS}}}\fi
\ifx \arxivurl  \undefined \def \arxivurl#1{\href{http://arxiv.org/abs/#1}{\textsf{arXiv}}}\fi

\def\arcsec{\hbox{$^{\prime\prime}$}}

\def\degns{\ifmmode^\circ\else$^\circ$\fi}
\def\deg{\ifmmode^\circ\else$^\circ$\fi}

\def\eg{{\it e.g.}~}
\def\egb{{\it e.g.}}

\def\ie{{\it i.e.}~}
\def\etal{{\it et~al.}~}

\def\gsim{\lower.4ex\hbox{$\;\buildrel >\over{\scriptstyle\sim}\;$}}
\def\lsim{\lower.4ex\hbox{$\;\buildrel <\over{\scriptstyle\sim}\;$}}

\begin{document}

\begin{article}

\begin{opening}

\title{Observing the Sun with the Atacama Large Millimeter-submillimeter
Array (ALMA): Fast-Scan Single-Dish Mapping}

%
\author[addressref={1},corref,email={stephen.white.24@us.af.mil}]{\inits{S.M.}\fnm{S.M.}~\lnm{White}}
\author[addressref={2}]{\inits{K.}\fnm{K.}~\lnm{Iwai}}
\author[addressref={3,4}]{\inits{N.M.}\fnm{N.M.}~\lnm{Phillips}}
\author[addressref={5}]{\inits{R.E.}\fnm{R.E.}~\lnm{Hills}}
\author[addressref={3,7}]{\inits{A.}\fnm{A.}~\lnm{Hirota}}
\author[addressref={6}]{\inits{P.}\fnm{P.}~\lnm{Yagoubov}}
\author[addressref={3,4}]{\inits{G.}\fnm{G.}~\lnm{Siringo}}
\author[addressref={7}]{\inits{M.}\fnm{M.}~\lnm{Shimojo}}
\author[addressref={8}]{\inits{T.S.}\fnm{T.S.}~\lnm{Bastian}}
\author[addressref={3,8}]{\inits{A.S.}\fnm{A.S.}~\lnm{Hales}}
\author[addressref={3,7}]{\inits{T.}\fnm{T.}~\lnm{Sawada}}
\author[addressref={3,7}]{\inits{S.}\fnm{S.}~\lnm{Asayama}}
\author[addressref={7}]{\inits{M.}\fnm{M.}~\lnm{Sugimoto}}
\author[addressref={9}]{\inits{R.G.}\fnm{R.G.}~\lnm{Marson}}
\author[addressref={7}]{\inits{W.}\fnm{W.}~\lnm{Kawasaki}}
\author[addressref={7}]{\inits{E.}\fnm{E.}~\lnm{Muller}}
\author[addressref={7}]{\inits{T.}\fnm{T.}~\lnm{Nakazato}}
\author[addressref={7}]{\inits{K.}\fnm{K.}~\lnm{Sugimoto}}
\author[addressref={10}]{\inits{R.}\fnm{R.}~\lnm{Braj\v{s}a}}
\author[addressref={11}]{\inits{I.}\fnm{I.}~\lnm{Skoki\'{c}}}
\author[addressref={11}]{\inits{M.}\fnm{M.}~\lnm{B\'{a}rta}}
\author[addressref={12}]{\inits{S.}\fnm{S.}~\lnm{Kim}}
\author[addressref={8}]{\inits{A.J.}\fnm{A.J.}~\lnm{Remijan}}
\author[addressref={3,4}]{\inits{I.}\fnm{I.}~\lnm{de~Gregorio}}
\author[addressref={3,8}]{\inits{S.A.}\fnm{S.A.}~\lnm{Corder}}
\author[addressref={13}]{\inits{H.S.}\fnm{H.S.}~\lnm{Hudson}}
\author[addressref={14,15,16}]{\inits{M.}\fnm{M.}~\lnm{Loukitcheva}}
\author[addressref={14}]{\inits{B.}\fnm{B.}~\lnm{Chen}}
\author[addressref={17}]{\inits{B.}\fnm{B.}~\lnm{De Pontieu}}
\author[addressref={14}]{\inits{G.D.}\fnm{G.D.}~\lnm{Fleishmann}}
\author[addressref={14}]{\inits{D.E.}\fnm{D.E.}~\lnm{Gary}}
\author[addressref={18}]{\inits{A.}\fnm{A.}~\lnm{Kobelski}}
\author[addressref={19}]{\inits{S.}\fnm{S.}~\lnm{Wedemeyer}}
\author[addressref={20}]{\inits{Y.}\fnm{Y.}~\lnm{Yan}}

\address[id=1]{Space Vehicles Directorate, Air Force Research
Laboratory, 3550 Aberdeen Avenue SE, Kirtland AFB, NM 87117-5776, USA}
\address[id=2]{National Institute of Information and Communications
Technology, Koganei 184-8795, Tokyo, Japan}
\address[id=3]{Joint ALMA Observatory (JAO), Alonso de C\'{o}rdova 3107, 
Vitacura 763-0355, Santiago, Chile}
\address[id=4]{European Southern Observatory, Alonso de C\'{o}rdova 3107, 
Vitacura 763-0355, Santiago, Chile}
\address[id=5]{Astrophysics Group, Cavendish Laboratory, JJ Thomson
Avenue, Cambridge CB3 0HE, UK}
\address[id=6]{European Southern Observatory (ESO),
Karl-Schwarzschild-Strasse 2, 85748 Garching bei M\"{u}nchen, Germany  }
\address[id=7]{National Astronomical Observatory of Japan (NAOJ), 2-21-1 Osawa,
Mitaka, Tokyo 181-8588, Japan}
\address[id=8]{National Radio Astronomy Observatory (NRAO), 520 Edgemont
Road, Charlottesville, VA 22903, USA}
\address[id=9]{National Radio Astronomy Observatory (NRAO), Pete V.
Domenici Science Operations Center, 1003 Lopezville Road, Socorro, NM 87801}
\address[id=10]{Hvar Observatory, Faculty of Geodesy, University of
Zagreb, Ka\v{c}i\'{c}eva 26, 10000 Zagreb, Croatia}
\address[id=11]{Astronomical Institute, Czech Academy of Sciences, Fri\v{c}ova
298, 251 65 Ond\v{r}ejov, Czech Republic}
\address[id=12]{Korea Astronomy and Space Science Institute, Daejeon
305-348, Korea}
\address[id=13]{School of Physics and Astronomy, University of Glasgow, 
Glasgow, G12 8QQ, Scotland, UK}
\address[id=14]{Center For Solar-Terrestrial Research, New Jersey 
Institute of Technology, Newark, NJ 07102, USA}
\address[id=15]{Max-Planck-Institut for Sonnensystemforschung, 
Justus-von-Liebig-Weg 3, 37077 G{\"o}ttingen, Germany}
\address[id=16]{Astronomical Institute, St. Petersburg University,
Universitetskii pr. 28, 198504, St. Petersburg, Russia}
\address[id=17]{Lockheed Martin Solar \& Astrophysics Lab, Org. A021S,
Bldg. 252, 3251 Hanover Street, Palo Alto, CA 94304, USA}
\address[id=18]{Center for Space Plasma and Aeronomic Research, The
University of Alabama Huntsville, Huntsville, AL 35899, USA}
\address[id=19]{Institute of Theoretical Astrophysics, University of Oslo, 
Postboks 1029 Blindern, N-0315 Oslo, Norway}
\address[id=20]{National Astronomical Observatories, Chinese Academy of
Sciences A20 Datun Road, Chaoyang District, Beijing 100012, China}

%
\runningauthor{S. M. White \etal}
\runningtitle{ALMA Solar Single-Dish Observations}

\begin{abstract}
The Atacama Large Millimeter-submillimeter Array (ALMA) radio telescope
has commenced science observations of the Sun starting in late
2016. Since the Sun is much larger than the field of view of individual
ALMA dishes, the ALMA interferometer is unable to measure the background
level of solar emission when observing the solar disk. The absolute
temperature scale is a critical measurement for much of ALMA solar
science, including the understanding of energy transfer through the
solar atmosphere, the properties of prominences, and the study of shock
heating in the chromosphere. In order to provide an absolute temperature
scale, ALMA solar observing will take advantage of the remarkable 
fast-scanning capabilities of the ALMA 12\,m dishes to make single-dish
maps of the full Sun. This article reports on the results of an extensive
commissioning effort to optimize the mapping procedure, and it describes the 
nature of the resulting data. Amplitude calibration is discussed in detail: 
a path that utilizes the two loads in the ALMA calibration system as well
as sky measurements is described and applied to commissioning data.
Inspection of a large number of single-dish datasets shows significant 
variation in the resulting temperatures, and based on the
temperature distributions we derive quiet-Sun values at disk
center of 7300 K at $\lambda\,=\,3$ mm  and 5900 K at $\lambda\,=\,1.3$ mm. 
These values have statistical uncertainties of order 100 K, but
systematic uncertainties in the temperature scale that may be
significantly larger.  Example images are
presented from two periods with very different levels of solar activity.
At a resolution of order 25\arcsec, the 1.3 mm wavelength images show
temperatures on the disk that vary over about a 2000 K range. Active
regions and plage are amongst the hotter features while a large
sunspot umbra shows up as a depression and filament channels are
relatively cool. Prominences above the solar limb
are a common feature of the single-dish images.
\end{abstract}

%

\end{opening}

\setlength\textheight{666pt}
\setlength\voffset{0mm}


\section{Introduction}

The earliest observations of the Sun at millimeter wavelengths were carried 
out with the single-dish telescopes used to develop observing at such
short wavelengths, frequently treating the Sun as a calibration source 
\citep[\egb][]{Sin52,WCM57,Coa58,ToS61}. Subsequently, more attention was
paid to understanding what the observations were telling us about the
solar atmosphere
\citep[\egb][]{NBL68,New69,BuT70,Kun70,LiH76,Lab78,KSS82,RiS76,UKH86,KIS86,CNT92a,VPU92,BEZ93a,BBT07}.
Millimeter wavelengths probe down into the chromosphere of the Sun where
the radiation is optically thick, in
contrast to longer radio wavelengths that tend to be dominated by bright
coronal sources \citep[\egb][]{Whi99a}. Radio emission at millimeter
wavelengths is in the Rayleigh-Jeans limit, 
such that the measured radio brightness temperature is the actual
electron temperature in the radiating layer when optically thick, 
and this makes such
observations an important aspect of understanding the temperature
structure of the solar atmosphere \citep[\egb][]{VAL76,WBB16}.

Single-dish observations of the Sun have been somewhat neglected in recent
years \citep[see however ][]{IwS15} because they have not kept up with the
improved spatial resolution available at non-radio wavelengths and with
radio--interferometer data: the spatial resolution of single-dish
observations is fixed by the aperture of the dish and the wavelength. 
In addition, the traditional (at least prior to the last decade) 
single-dish mapping method
consisting of dwells at a raster of grid points is a slow 
technique, particularly with large-aperture dishes constrained by their
weight to move slowly, and it is thus poorly suited to tracking the dynamic 
phenomena typical of the solar chromosphere. 

However, with the opening of solar observations by the Atacama Large
Millimeter-submillimeter Array \citep[ALMA:][]{WoT09}, single-dish data are expected
again to be an important feature of studies of the solar chromosphere
\citep{PHB15}.  ALMA will provide superb high-spatial-resolution images of 
the chromosphere with its large interferometric array operating at
wavelengths from 3 to 0.3 mm \citep[see][]{SBH17}. However, the field of view 
of the interferometric observations is relatively small, and the
interferometer is unable to measure spatial scales larger than the
fringe spacing corresponding to the shortest baseline in the array, 
making it insensitive to the absolute
temperature of the solar atmosphere. The absolute temperature scale, and
temperature variation with frequency corresponding to different heights
in the chromosphere, is crucial for a wide range of ALMA solar
science \citep[\egb][]{WBB16}. Single-dish data can restore
information on the absolute temperature scale missing from the
interferometer data. For mapping of larger areas in which the
interferometer carries out a mosaic (multiple pointings), the
single-dish image will be used to fill in the large spatial scales during
deconvolution \citep[\egb][]{EkR79,KSS09}. 
Two other developments make this technique
attractive: in recent years the mapping method for single-dish data has
changed to exploit continuous movement of the dish \citep[the so-called
``on-the-fly'' technique;][]{MEG07}, making it much faster; 
and ALMA incorporates 12\,m dishes
specifically designed with the ability to scan rapidly yet precisely in order 
to exploit this technique. A major advantage of
these observations is the speed with which the Sun can be mapped, which
both permits the study of time-varying phenomena and 
minimizes the impact of variation in terrestrial atmospheric transmission,
which is a major issue for short-wavelength millimeter 
observations.\footnote{Fast-scanning observing with the ALMA dishes was 
initially
developed by Richard Hills and Neil Phillips, with non-solar observing
in mind, during several commissioning campaigns starting in 2010. These
efforts are documented in the ALMA Commissioning and Science
Verification task CSV-203 and attached sub-tasks; specific solar
developments are documented under CSV-3162 (2014 campaign) and
CSV-3244 (2015).}

It is planned that ALMA interferometer observations of the Sun will be 
complemented by single-dish (SD) data from several 12\,m 
antennas, permitting redundancy in case of failures at one or
more of the antennas.  Initially, ALMA solar single-dish
mapping will be monomode: full-disk total-power observations with fixed 
parameters will be carried out continuously at the interferometer observing
frequencies.  During a solar commissioning campaign in December 2015, tests 
were carried out in order to have a basis for recommending the SD observing
conditions for full science observations of the Sun. These started in
late 2016 in Bands 3 (84\,--\,116 GHz) and 6 (211\,--\,275 GHz).  This article
describes the results of the tests, the properties of the SD data, its
calibration, and it discusses initial results on the temperatures of the layers
that ALMA will probe. 

\begin{figure}[t]
\epsfxsize=120mm
\centerline{\epsffile{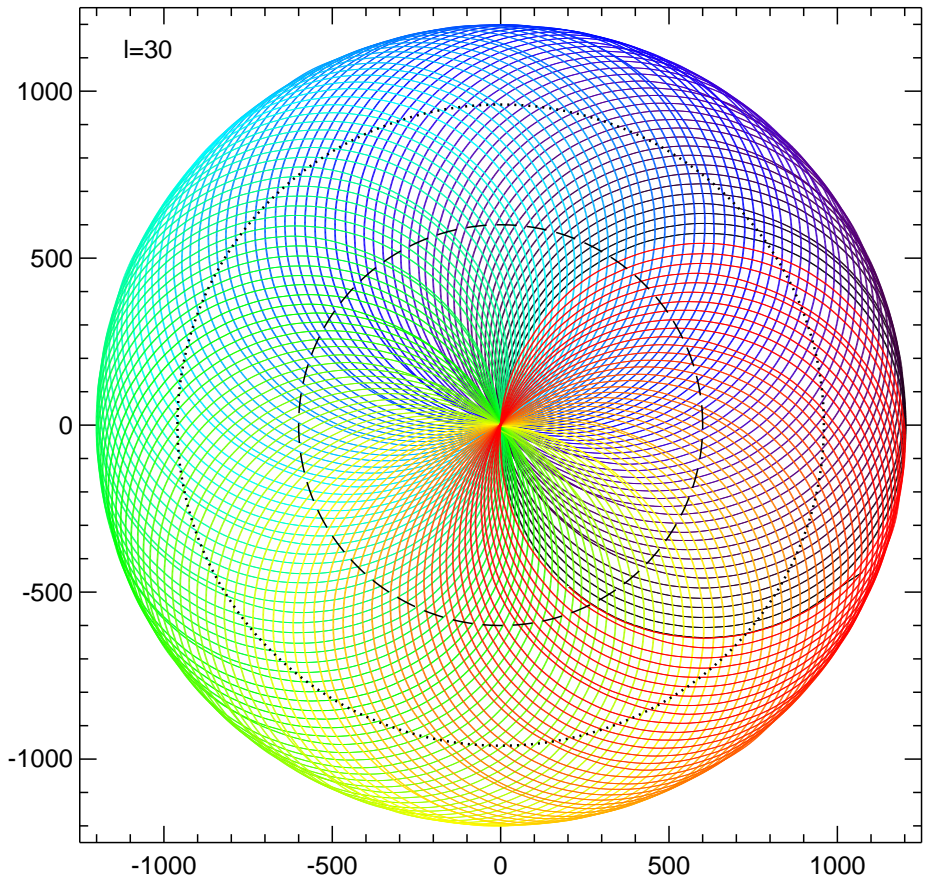}}
\caption{The double-circle pattern for a 2400\arcsec\ diameter field of
view with ``map spacing'' of 30\arcsec. The dashed-line circle at a
radius of 600\arcsec\ is the ``major circle'' track of the centers of the
individual minor circles. The dotted line is the (average) 
solar limb at a radius of 960\arcsec.
The color variation of the telescope track is simply a rainbow color table
intended to show the progress of the pointing location with time.  Mapping
started and ended at about the ``3 o'clock'' location on the figure.
}
\label{fig:double}
\end{figure}

\section{Fast-Scan Imaging With ALMA 12\,m Dishes}

ALMA fast-scan solar imaging will primarily be carried out with four 12\,m
dishes constructed by the Mitsubishi Electric Corporation
\citep[MELCO:][]{IMS09,SKI09} and labelled PM01-PM04.\footnote{The 
twenty-five 12\,m antennas built for ALMA by the AEM Consortium, 
normally used for interferometer observations, are also direct-drive 
antennas that perform well as fast-scanning dishes, but they
will not be discussed in this article.} 
In order to optimize the ability of the dishes to maintain
pointing accurately as they scan, an initial characterization of the
individual antenna servo response is carried out by comparing commanded
and reported antenna positions under a range of conditions and
determining the servo gain. Corrections (usually small) can then be 
implemented in the drive software \citep[\egb][]{Hil16}. 
During the initial phase of ALMA solar observations, ALMA will
offer only full-disk SD observations. This will consist of
a circular field of view of diameter 2400\arcsec\ that will be mapped using a
``double-circle'' scanning pattern.
This provides a region of over 200\arcsec\ off the solar limb
encompassing solar prominences above the limb as well as 
blank-sky regions for checks on calibration.  Since there will be no 
opportunity during the initial solar observations to change the interferometer 
pointing location on the
timescales of minutes that would be necessary to track, \eg a prominence
eruption, this field of view should address all likely science use cases 
relevant for initial observations, while simultaneously
optimizing the cadence of single-dish mapping (which is the cadence at
which zero-spacing information is obtained for the interferometer
target field of view). These restrictions on SD observing may be relaxed in 
future observing cycles.

Figure \ref{fig:double} shows the scanning pattern on the sky for a double-circle
fast-scanning map, taken from actual data from PM02 during the December 2015
commissioning campaign. Essentially the mapping pattern consists of (``minor'')
circles with
diameter half that of the overall field of view (FOV) whose centers move
steadily in a (``major'') circle around the center of the field of view
at a distance that is
one-quarter of the diameter of the field of view. An advantage of this
pattern is that the continual circular motion allows a steady velocity
to be maintained, whereas other patterns, such as
Lissajous, often require sharp turns with higher acceleration.
Every minor circle passes
through the center of the field of view, which means that this location is
oversampled compared to regions halfway out to the edge of the FOV, but
has the advantage that fluctuations in terrestrial 
atmospheric transmission (which can
have a major impact on the received flux and thus produce large artefacts
in the resulting maps) can simply be tracked and removed by requiring the
central brightness temperature to remain stable. This is not likely to be
important for Bands 3 and 6, but it will matter more for later observations at
shorter wavelengths where atmospheric opacity is larger.

\begin{figure}[t]
\epsfxsize=160mm
\centerline{\epsffile{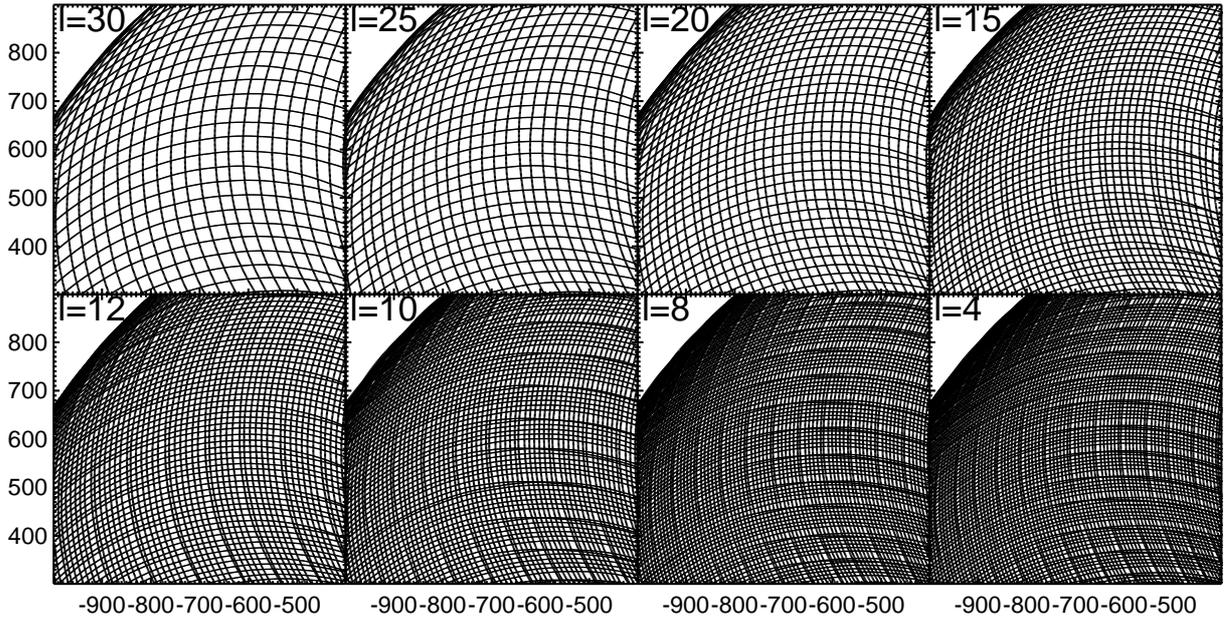}}
\caption{Portions of the fast-scanning patterns for different values of
the {\textsf{samplingLength}} parameter \textsf{l}. Each 1-millisecond sample is plotted
as a point, but since the separation of samples along each minor circle is
1.5\arcsec, they appear to overlap on the plot. The region plotted is
around the major circle at radius 600\arcsec\ from disk center where we
expect the largest holes in the sampling pattern to occur.
}
\label{fig:pattern}
\end{figure}

The SD observing program has a parameter {\textsf{samplingLength}} [\textsf{l}] which
controls the number of minor circles in a pattern by changing the spacing
of the centers of the minor circles as they move around the major circle.
This parameter needs to be smaller at higher frequencies in order to match
the observing pattern to the higher spatial resolution of the dishes.
A second parameter, {\textsf{subscanDuration}}, specifies the duration of on-source
mapping, and these two numbers have to be coordinated in order to ensure
that a complete map is achieved. In addition, there is no value to
having the {\textsf{subscanDuration}} be significantly longer than required
for a full pattern because the additional partial coverage of the solar
disk will take extra time without enhancing the resulting image, so the
{\textsf{subscanDuration}} is specified to match the duration of a
full pattern.

\begin{figure}[t]
\epsfxsize=120mm
\centerline{\epsffile{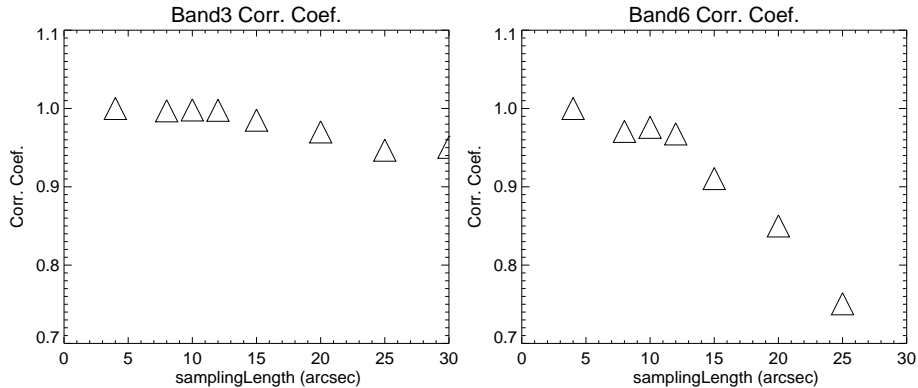}}
\caption{Correlation coefficient of intensity as a function of
\textsf{samplingLength} \textsf{l} for Bands 3 and 6, from \citet{Iwa16}. The plot shows a
significant decline for Band 3 at \textsf{l}$\,>\,20$\arcsec\ and for Band 6 at
\textsf{l}$\,>\,12$\arcsec.
}
\label{fig:sims}
\end{figure}

There are limits to both the permitted
maximum acceleration [3\deg\ s$^{-2}$] and velocity [1\deg\ s$^{-1}$] that
can safely be used with the PM antennas, 
and these are taken into account by the control software when converting the
desired pointing pattern into the sinusoids to be commanded to the antenna 
drives, effectively scaling down the frequencies as needed. In the case of the
large maps needed for the solar disk, 
the acceleration limit determines the minor-circle frequency
to be about 0.675 Hz and velocity 0.71\deg\ s$^{-1}$.
Tests were carried out with the PM antennas to determine
the durations of a full pattern for likely values of \textsf{l}, and
the results are shown in Table \ref{tab:patterns}.

\begin{table}[h]
\caption{Pattern durations as a function of {\textsf{samplingLength}} for a
2400\arcsec-diameter circular field-of-view.}
\begin{tabular}{ccc}
 & \\
\hline
\textsf{samplingLength} & Pattern duration & Minor circles \\ 

 [arcsec] & [seconds] & per pattern \\
\hline
30 & 187 & 125 \\
25 & 224 & 150 \\
20 & 280 & 188 \\
15 & 373 & 251 \\
12 & 467 & 314 \\
10 & 560 & 377 \\
8 & 701 & 471 \\
4 & 1397 & 943 \\
\hline
\end{tabular}
\label{tab:patterns}
\end{table}

The results are consistent with a linear relationship between 
\textsf{samplingLength} and pattern duration resulting from the constant scan
rate. The region with the largest spacing (on the sky) between samples is
along the major circle. The number of minor circles (each of length 
3770\arcsec\ for
a 600\arcsec\ radius) essentially matches the length of the major circle (also
600\arcsec\ radius and 3770\arcsec\ long) divided by the 
\textsf{samplingLength} parameter. Since
each minor circle crosses the major circle twice, on average the largest
distance between sampled points is half the {\textsf{samplingLength}} parameter, 
barring unfortunate cases where the minor circle crossings fall on top of each
other.
In practice, the double-circle
pattern seen in fast-scanning observations is not exactly regular. This
can be seen by careful inspection of Figure~\ref{fig:double}, and Figure
\ref{fig:pattern} shows more detail for each of the eight \textsf{l}-values
listed in Table \ref{tab:patterns}.
Irregularities in the tracks are clearly visible in all patterns,
presumably due to imperfections in the servo responses,
but the scale of the irregularities in the patterns is not large 
enough to prevent successful mapping (which grids the sampled points on
the sky according to the actual measured pointing locations returned by
encoders on each antenna).

Presently there is significant overhead (of order nine minutes) for each single-dish map:
both focus and pointing adjustments are first carried out using a very bright 
source such as a large planet, and atmospheric calibration scans are
carried out before and after the solar data acquisition.
The nominal spatial resolution of a single dish (with the
beam response tapered by 10 dB at the edges to minimize the effects of
rear spillover, which is a common design practice for such telescopes)
is 1.13$\lambda/D$ (full width at half maximum, or FWHM),
where $\lambda$ is the observing
wavelength and $D$ is the dish diameter. For the 12\,m PM antennas, this
corresponds to 58\arcsec\ at 100 GHz in Band 3 and 25\arcsec\ at 230 GHz
in Band 6.
The choice of optimal {\textsf{samplingLength}} is a trade-off between adequate
sampling of the image plane and the additional time required for finer sampling.
Nyquist sampling of the single dish images
requires sampling distances of at most 30\arcsec\ in Band 3 and
12\arcsec\ in Band 6.  \citet{Iwa16} has
investigated the effect of \textsf{l} on image quality. He took a model solar
image, convolved it with the telescope response, sampled it per
the fast-scanning patterns shown above with a range of values of \textsf{l}
and then imaged the result
using a standard triangulation gridding method. He then took a torus just
inside the solar limb and determined correlation coefficients against
the most densely sampled pattern (\textsf{l}$\,=\,4$\arcsec) as a
function of \textsf{l}, with the assumption that sampling with \textsf{l} much smaller
than the primary beam provides an
appropriate reference curve (unity correlation).
The results are shown in Figure \ref{fig:sims}. The plot shows a
significant decline in the correlation coefficients beyond
\textsf{l}$\,=\,20$\arcsec\ for Band 3 and \textsf{l}$\,=\,12$\arcsec\ for Band 6.
Based on these results, the fast-scan SD mapping currently uses 
{\textsf{samplingLength}s} of \textsf{l}$=20$\arcsec\ for Band 3 (total duration of 13 
minutes per full-disk execution including the calibrations described above) 
and \textsf{l}$=10$\arcsec\ for Band 6 (17 minutes per execution). These parameters
achieve Nyquist sampling in the solar disk images.

\begin{figure}[t]
\epsfxsize=120mm
\centerline{\epsffile{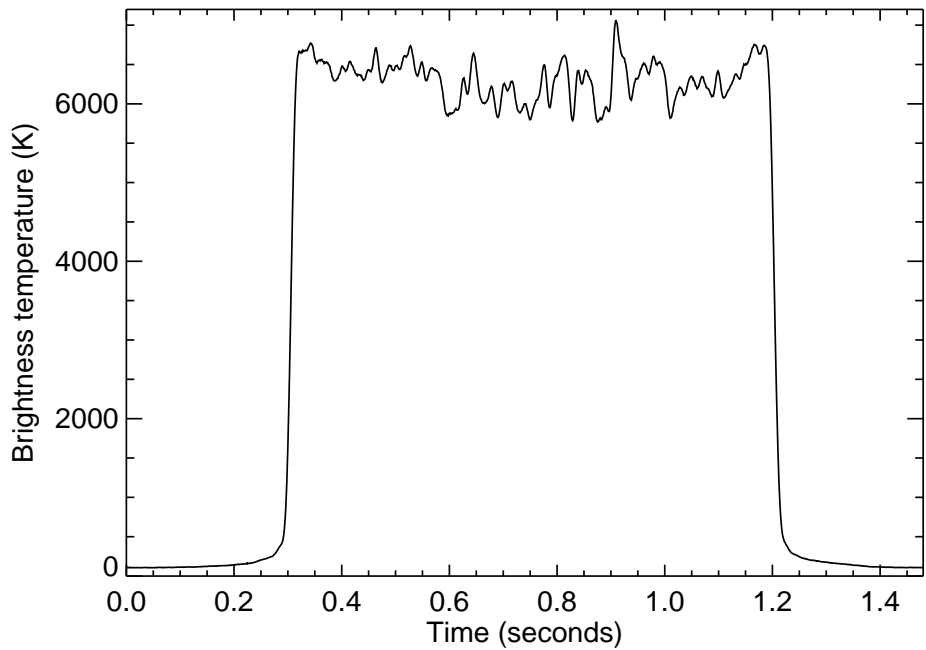}}
\caption{The TP amplitude variation around a single circle
of the double-circle pattern. The plot consists of 1480 individual
one-millisecond TP measurements at 230 GHz, calibrated to brightness temperature.
The 600\arcsec\ radius circle starts off the
limb at 1200\arcsec\ from disk center and passes onto the limb: from the
low level of fluctuations at off-limb positions it is clear that the
noise level is insignificant in the one-millisecond integrations and the
Sun's signal dominates each sample. All of the temporal variations seen on the disk
represent the telescope sweeping over real structures in the solar
atmosphere.
}
\label{fig:onecircle}
\end{figure}

\section{Sampling Time}

In addition to the spatial sampling of the scanning pattern embedded in
the {\textsf{samplingLength}} parameter, the quality of the resulting image
depends on the interval between data samples. The data being used
for SD mapping come from the total power (TP) detectors for each of the
four 2 GHz-wide basebands (equivalent to spectral windows for our purposes) 
provided by the two-stage heterodyne conversion employed by ALMA
\citep[described further by ][]{SBH17}. These are simple square-law 
detectors with a hardware sampling time of 0.5 milliseconds.
Test fast-scanning observations have generally been carried out with TP
data recording times of one or two milliseconds: the Sun is so bright that a 
one-millisecond sample on the disk is dominated by the solar signal.  
This is demonstrated in Figure \ref{fig:onecircle}, which shows 1480 raw
one-millisecond power measurements for a single minor circle of the
double-circle pattern, passing onto and then off the solar disk. 
The temporal variation seen in this figure is due to real structure along the 
path of the antenna across the solar disk: the noise level for a system
temperature of order 1000 K (see below) with 2 GHz bandwidth in 0.5
milliseconds
is of order 1 K, \ie $<10^{-3}$ of the received power in each IF band and 
polarization, and small
compared to the point-to-point variations in amplitude.

For a 2400\arcsec\ double-circle
pattern the average velocity of the telescope of 
0.7\deg\ s$^{-1}$ (independent of the {\textsf{samplingLength}} parameter)
results in a spatial sampling rate of about 2.5\arcsec\ on the sky for 
a TP sampling time of one millisecond.  Each
minor circle takes about 1.5 seconds (\ie
the telescope field of view passes through the
disk center region every 1.5 seconds). Such a fine spatial sampling
along the path is not
required for Nyquist sampling in Bands 3 and 6, but short integrations are
valuable for providing flexibility in removing any bad measurements.
For the current solar SD observations in Bands 3 and 6, the TP sampling time is 
one millisecond.

The telescope-pointing information coming from the drive software 
is not sampled on the same timescale:
it has a fixed 48-millisecond interval. During calibration, the ALMA mapping
software has to interpolate the pointing data onto the one-millisecond timescale
of the total-power samples. The interpolation is not linear, since the
telescope path on the sky is curved, so several pointing measurements
must be used to get the correct path.

The resulting single-dish files are small by ALMA interferometer standards, 
but still of significant size: a Band 3 dataset in standard ALMA format with a 
{\textsf{samplingLength}} of 20\arcsec\ and four antennas is 292 MB.  For Band 6, 
a file with {\textsf{samplingLength}} of 10\arcsec\ and four antennas is 532 MB.

\section{Gridding of Fast-Scan Data}

In addition to the dish aperture and tapering of the antenna response to
minimize spill-over, the effective spatial resolution of an on-the-fly 
single-dish image
depends on the manner in which the data samples are gridded into an
image: the data are smoothed at the irregularly-spaced coordinates 
corresponding to the pointing locations of the dish when each sample was
acquired, and then they are interpolated onto a regular grid. 
The process of interpolation effectively acts as a convolution, and 
\citet{MEG07} discuss this process and the range of functions typically 
used. The
two gridding functions commonly employed in ALMA single-dish imaging are
a spheroidal function (``SF'') and a ``GJinc'' function (Gaussian multiplied 
by a Bessel function). These are implemented in the single-dish mapping
tasks in the software package {\it Common Astronomy Software Applications} 
(\textsf{CASA}, see \textsf{casa.nrao.edu}) that provides the
standard analysis path for ALMA data. \textsf{CASA} also provides tasks to carry
out amplitude calibration, discussed further below. \textsf{CASA} scripts for the
end-to-end processing of ALMA single-dish fast-scan data are
provided with the solar Science Verification data release (see
\textsf{almascience.nrao.edu/alma-data/science-verification}).

\citet{BrH14} have analyzed the effect of
different gridding choices on ALMA single-dish data and recommend SF
gridding as the standard method for ALMA mapping, while the 
GJinc function may give slightly better spatial resolution. 
We note that processed images may still show trace artefacts of the original
double-circle scanning pattern, particularly at the limb. Small errors
in the assumed pointing of the dish (or small errors in the assumed
timing of the samples, which has the same effect)
can have a large effect at the limb
because the large temperature difference between the cold sky and hot
solar disk magnifies the error when the actual relative proportions of
those two contributions to the measured power do not match the
proportions appropriate to the assumed pointing.

\section{Elevation Limitations}

The ALMA antennas have azimuth-elevation mounts, which means that at high
elevations the azimuth drives must scan much faster than at low elevations
for the same distance moved on the sky. Because fast scanning is already
driving the antennas at close to the upper limits of their slewing rates,
it cannot be carried out at high elevations\footnote{The solar team
can confirm that trying to observe at higher elevations does drive the
antennas into a non-functioning state: the
prolonged excessive accelerations can cause the drive power amplifiers
of the PM antennas to trigger an over-current cut-out, requiring
manual reset by a human at the antenna to recover operation.}: 
empirically, the solar fast-scan patterns described here have been found
to be usable up to a maximum elevation of 70\deg, corresponding to an
azimuth acceleration of up to 8.8 deg s$^{-2}$.
The slew rates required for tracking a single point on the solar surface
are much less demanding, and interferometric observing is permitted to 
higher elevations (typically recommended up to about 82\deg), limited by
the potential of azimuth slew durations to and from the phase calibrator 
becoming excessive, and the risk of either the science target or 
phase calibrator entering a small zenith
avoidance zone that will cause the observation to abort. There is also
a lower limit to the elevation of solar interferometric observations of
about 40\deg, imposed by mutual shadowing at lower elevations of the 
7\,m dishes in the compact array.  The implications of the elevation limit on 
single-dish mapping depend on the time of year that solar observations take 
place.  Since ALMA is at a latitude of 23\deg\ south, the Sun passes directly
overhead at local noon at the December solstice, and the upper elevation limit
means that fast-scan observing is
not permitted for a three-hour period centered on local noon. 
At other times of the year when the Sun does not reach such
high elevations, the blocked period will be shorter, but it should be
recognized that this restriction on the SD mapping can have an impact on
the scheduling of solar programs.

\section{Receiver Set-Up}

As described by \citet{SBH17}, observations of the bright solar
flux with ALMA require
special conditions to achieve a linear response in the sensitive receivers.
Originally it was planned that special solar attenuators would be placed
in the beam path to reduce the signal reaching the receivers, but this
approach has a number of disadvantages: notably the amount of attenuation is
fixed,  and the attenuators will impose complex gain variations on the
signals that would need careful and extensive measurements at every
observing frequency on every antenna for correction. Furthermore, since the solar
attenuators are located in the same filter-wheel as the calibration
loads, standard ALMA calibration cannot be carried out through the solar
attenuators. Subsequently, 
\citet{Yag13} described a detuning technique in which the ALMA
Superconductor-Insulator-Superconductor (SIS) 
mixer voltages  are not biassed to their usual, and highest gain,
values near the middle of the first photon step below the superconducting
gap voltage, but to other photon steps above or below the gap voltage
that have lower gain. In addition, the local oscillator power supplied
to the SIS mixers can be used to fine-tune the resulting gain. 
These alternate ``solar tuning'' mixer-bias voltages have been
determined for the ALMA receivers in Bands 3 and 6 \citep[see further
discussion in][]{SBH17}. For single--dish solar observations, data will be 
taken with the ``MD2'' bias, \ie the second photon step above the superconducting
gap voltage for Band 3 and the first photon step above the gap for Band 6, 
resulting in system temperatures of order 1000 K in both bands, 
rather than the values of order 40 K resulting from normal bias: the
lower gain of the MD2 bias provides linearity for the wide range of solar
brightness temperatures (up to 8000 K) given the dynamic
range of the ALMA system. The receivers are
believed to respond linearly in this temperature range, although we
cannot rule out the possibility that there is a small amount of gain
compression \citep[a few percent: see][]{Iwa16b}; 
the linearity of the receivers with the MD2 bias at
the much higher antenna temperatures that may occur during solar flares
has not yet been investigated.

\begin{table}[h]
\caption{Continuum frequencies for ALMA Cyle 4 solar observations. These
are the central frequencies of the 2-GHz-wide basebands.}
\begin{tabular}{lcccc}
\hline
       &  BB 1 &  BB 2 &  BB 3 &  BB 4 \\
       &  GHz  &  GHz  &  GHz  &  GHz  \\
\hline
Band 3 & \ 93.0 & \ 95.0 & 105.0 & 107.0 \\
Band 6 & 230.0 & 232.0 & 246.0 & 248.0 \\
\hline
\end{tabular}
\label{tab:freqs}
\end{table}

The observing frequencies used for single-dish mapping will be 
fixed to match the interferometer observations \citep[see ][]{SBH17}: 
there will be four continuum basebands, each 2 GHz wide.
The frequencies are shown in Table \ref{tab:freqs}.
The single-dish mapping will initially measure continuum emission only
and will be carried out using 
the total power detectors associated with each baseband in both linear
polarizations (X and Y) simultaneously, providing eight different data
streams. In the future, spectral-line
capability will be available for solar observations: at present,
single-dish spectral-line observations require the use of a correlator.

\section{Calibration Measurements}
\label{sec:calmeas}

Since we are using the ALMA single-dish data to supply the background
brightness temperature scale that the interferometer data cannot
measure, calibration is a very important aspect of the solar
fast-scanning observations. A standard calibration technique 
is to use observations of a reference source of known flux to scale
the data: this technique is difficult for the solar observations due
to the high system temperature and the lack of bright sources other than
the Moon that fill the beam as the Sun does.
Instead, we use an approach developed for
single-dish data, modified for our observing technique. Calibration
relies on measurements made with the ALMA Calibration Device
\citep{CMB08,YMW11,KCL12} located in each antenna: this system is capable of 
placing two microwave absorbers at different temperatures, 
the ``ambient'' (nominally 20\deg\ C) and ``hot'' (nominally 85\deg\ C) loads, 
in front of the receiver feed horns of any of the bands. At the beginning
and end of the SD observation, five-second measurements of the receiver signal
level are made for the following targets:

\begin{itemize}

\item a ``\textsf{sky}'' observation offset (by typically 2\deg) from
and at the same elevation as the target;

\item an ``\textsf{ambient}'' load observation that fills the beam path
at the temperature  of the thermally-controlled receiver cabin
(nominally 20\deg\ C);

\item a ``\textsf{hot}'' load observation that fills the beam path; 

\item a ``\textsf{zero}'' level measurement, which reports the power levels in
the detectors when no signal power is being supplied.
In the case of ALMA, the zero levels are very
stable over many days for a given antenna and receiver;

\item and finally, a second measurement of the power level on the sky,
called the ``\textsf{off}'' level,
made after the telescope has moved to the target (in our case, the Sun)
where the intermediate-frequency-stage (IF) attenuation levels
are reset to values appropriate for the power being seen. Since the Sun is 
much brighter than other sources, the increased attenuation results in a
lower power than the first sky measurement.

\end{itemize}

\noindent The fact that two different intermediate-frequency-stage 
attenuation settings are required means that, unlike normal single-dish
observations, two different receiver gains must be solved for.

Since the sky is being used as a calibration source, we need
to know its effective temperature contribution, which requires knowledge of 
the atmospheric temperature and the opacity of the atmosphere. These are
measured by a separate device, the water vapor radiometer (WVR), 
attached to each
antenna \citep{Hil10,NBG13}. This device measures the shape and intensity of the
water vapor line at 183 GHz, and in combination with an atmospheric model
\citep[in the case of ALMA, the ATM model: ][]{PCS01} provides the
precipitable water vapor (PWV) level that is an important opacity source at
millimeter wavelengths. The atmospheric model also includes the
contributions of other atmospheric species.

\section{Calibration Analysis}
\label{sec:calanal}

We label the median powers from the calibration scans
as $P_{\rm sky}$, $P_{\rm amb}$, $P_{\rm hot}$,
$P_{\rm zero}$, and $P_{\rm off}$ for sky, ambient, hot, zero, and off,
respectively; $T$ denotes the corresponding temperatures. 
The standard method for calibrating millimeter data is to determine the
system temperature [$T_{\rm sys}$], \ie the total thermal noise level in the
system excluding the target source, and then apply it to the power
measurements according to some form of the algorithm

\begin{equation}
T_{\rm src}\ =\ T_{\rm sys}\ {{P_{\rm src}\,-\,P_{\rm ref}} \over P_{\rm ref}}
\label{e_basic}
\end{equation}

\noindent \citep[\egb][]{UlH76,KuU81,Man93} where $P_{\rm src}$ is the power 
measured while pointing at the target source, and $P_{\rm ref}$ 
is the power measured at a reference position that corresponds to the 
$T_{\rm sys}$ measurement (\ie no source power). Calibration by this method
amounts to determining $T_{\rm sys}$. 

\subsection{Dual-Load Calibration}

An approach that uses both loads in the ALMA Calibration System
is as follows. The power measured by a sky observation can be expressed as

\begin{equation} 
P_{\rm sky}\ =\ G\{ T_{\rm rec}\,+\,(1-{{\rm e}^{-\tau}})\eta_{\rm l}
T_{\rm atm}\,+\,(1\,-\,\eta_{\rm l})\,T_{\rm spill}\,+\,{{\rm
e}^{-\tau}}\eta_{\rm l} T_{\rm CMB} \}\,+\,P_{\rm zero}
\end{equation}

\noindent where $G$ is the gain; $T_{\rm rec}$ is the intrinsic thermal
noise of the receiver;
$\eta_{\rm l}$ is the antenna efficiency for a source that is much larger than
the primary beam;
$T_{\rm atm}$ is the temperature of the (optically thin) layer of the
atmosphere
that dominates the optical depth; and $\tau$ is the opacity of the
atmosphere (for a given telescope elevation $El$, $\tau$
is the zenith opacity $\tau_0$ times the ``air
mass'' $1/\sin(El)$). $\eta_{\rm l}$ is referred to as the {\it forward
efficiency} \citep{TaS12}, and it is the product of the efficiency
$\eta_{\rm r}$ that accounts for 
ohmic losses (heating of the dish surface by the incident radiation) and the
efficiency $\eta_{\rm rss}$, the {\it rear spillover efficiency}, that accounts for
signal not reflected from the main dish 
(\eg ground illuminated beyond the edge of the dish)
that reflects off the subreflector or other surface above the
dish and enters the receiver. The effective temperature of this
contribution [$T_{\rm spill}$] is usually at the ambient temperature of the
telescope. 

Above the atmosphere, the sky is filled
by the cosmic micro\-wave back\-ground (CMB).
The combination ${{\rm e}^{-\tau}} T_{\rm CMB}\,+\,(1-{{\rm e}^{-\tau}})
T_{\rm atm}$ in Equation (2) 
is the result of radiative transfer of the sky background
temperature $T_{\rm CMB}$ through a layer with optical depth $\tau$ and
temperature $T_{\rm atm}$.
Although we use simple temperatures in the formulae presented here, in each
case the relevant value to use is the equivalent Rayleigh-Jeans
radiation temperature at the point on the black-body curve 
corresponding to the observing frequency $\nu$,
$J(\nu,T)\,=\,(h \nu / k_{\rm B})/(\exp(h \nu / k_{\rm B} T)-1)$, where $h$ is
Planck's constant and $k_{\rm B}$ is Boltzmann's constant. For example,
the temperature of the CMB is 2.73 K, but the temperature we need
at millimeter wavelengths [$T_{\rm CMB}$] is the Rayleigh-Jeans
curve equivalent $J(\nu,2.73)\,\approx\,1$ K at 3 mm wavelength.
In the 3 mm window $J(T)\,\approx\,T-2.4$ to quite a good approximation
above a few tens of K, and $J(T)\,\approx\,T-5.5$ in the 1.3 mm window, so the
difference between $T$ and $J(T)$ is negligible for our purposes.

When the ambient and hot loads are placed in front of the receiver, they fill 
the optical path. The received powers are

\begin{equation} 
P_{\rm amb}\ =\ G\{ T_{\rm rec}\,+\,T_{\rm amb}\}\,+\,P_{\rm zero}
\end{equation}

\noindent and

\begin{equation} 
P_{\rm hot}\ =\ G\{ T_{\rm rec}\,+\,T_{\rm hot}\}\,+\,P_{\rm zero}.
\end{equation}

\noindent The power on the target source is

\begin{equation} 
P_{\rm src}\ =\ G_{\rm src}\{ T_{\rm rec}\,+\,{{\rm e}^{-\tau}}\eta_{\rm l}
\,T^*_{\rm src}\,+\,(1-{{\rm e}^{-\tau}})\eta_{\rm l}
T_{\rm atm}\,+\,(1\,-\,\eta_{\rm l})\,T_{\rm spill}\}\,+\,P_{\rm zero}
\end{equation}

\noindent where the gain [$G_{\rm src}$] is different from $G$ due to the
different IF attenuation setting used for the Sun. 
$T^*_{\rm src}$ is the source brightness
temperature. In the case of a source much larger than the telescope
beam, such as the Sun or the Moon, the target replaces the CMB as the 
background temperature. 
The power measured at the ``off'' position on the sky with the same
attenuation as the source measurement is, by analogy with Equation (2),

\begin{equation} 
P_{\rm off}\ =\ G_{\rm src}\{ T_{\rm rec}\,+\,(1-{{\rm e}^{-\tau}})\eta_{\rm l}
T_{\rm atm}\,+\,(1\,-\,\eta_{\rm l})\,T_{\rm spill}\,+\,{{\rm
e}^{-\tau}}\eta_{\rm l} T_{\rm CMB} \}\,+\,P_{\rm zero}
\end{equation}

We can now solve these equations for $T^*_{\rm src}$. The gain
$G$ is derived from Equations (3) and (4) as

\begin{equation} 
G\ =\ {{P_{\rm hot}\,-\,P_{\rm amb}} \over {T_{\rm hot}\,-\,T_{\rm amb}}}
\end{equation}

\noindent where typically $T_{\rm amb}\,=\,291$\deg\ K and $T_{\rm hot}\,=\,357$\deg\ K.
We can derive the source gain simply from the ratio of Equation (6) to
Equation (2),

\begin{equation} 
G_{\rm src}\ =\ G\,{{P_{\rm off}\,-\,P_{\rm zero}} \over {P_{\rm
sky}\,-\,P_{\rm zero}}}
\end{equation}

\noindent Note that when observing the Sun in MD2 mode with large system
temperature, $P_{\rm zero}$ is generally much smaller than the other
calibration measurements, but with normal mixer bias the IF attenuation is 
set to much higher values to ensure acceptable power levels from the Sun, 
and then $P_{\rm off}$, also measured with the higher attenuation, becomes much 
smaller and can be of the same order as $P_{\rm zero}$.  In that case
neglecting $P_{\rm zero}$ leads to major errors. Even with MD2 bias, as
we show below, neglect of $P_{\rm zero}$ can lead to errors of over 10\,\%
which are larger than we want for solar calibration. 
Note that $P_{\rm zero}$ is only relevant for data from the
total-power square-law detectors, not for correlation data.

The receiver temperature is determined from Equations (3) and (7) to be

\begin{equation} 
T_{\rm rec}\ =\
{{T_{\rm hot}\,(P_{\rm amb}\,-\,P_{\rm zero})\,-\,T_{\rm amb}\,(P_{\rm
hot}\,-\,P_{\rm zero})}
\over {P_{\rm hot}\,-\,P_{\rm amb}}}
\end{equation}

\noindent $T_{\rm atm}$ and $\tau$ are provided by the standard ALMA calibration
method as described above.  In any case, we have
sufficient measurements to determine the sky contribution 
without using the atmospheric data from the WVR: using Equations (2) and (7), 
we find, setting $P_{\rm sky}\,-\,P_{\rm zero}\ =\
G(T_{\rm rec}\,+\,T_{\rm sky})$, that the ``sky'' temperature contribution
(including rear spillover, which is actually not from the sky) is of the form

\begin{equation} 
T_{\rm sky}\ =\ (1-{{\rm e}^{-\tau}})\eta_{\rm l}\,T_{\rm atm}\,+\,{{\rm e}^{-\tau}}\eta_{\rm l}
T_{\rm CMB}\,+\,(1\,-\,\eta_{\rm l})\,T_{\rm spill}\ =\ {{P_{\rm
sky}\,-\,P_{\rm zero}} \over G}\,-\,T_{\rm rec}
\end{equation}

\noindent where all quantities on the right result from calibration
power measurements. Note that conventionally the system temperature (at the
telescope, not above the atmosphere) is $T_{\rm sys}\ =\ T_{\rm
rec}\,+\,T_{\rm sky}$.
We now subtract Equation (6) from Equation (5) (ignoring the CMB contribution of less than
1 K compared to the $\approx$1000 K system temperature) 
and apply Equations (7) and (8) to find

\begin{equation} 
T^*_{\rm src}\ =\ {{P_{\rm sky}\,-\,P_{\rm zero}} \over {P_{\rm
off}\,-\,P_{\rm zero}}}\
{{P_{\rm src}\,-\,P_{\rm off}} \over {P_{\rm hot}\,-\,P_{\rm amb}}} 
\ {{T_{\rm hot}\,-\,T_{\rm amb}} \over {\eta_{\rm l}\,{\rm e}^{-\tau}}}
\end{equation}

\noindent As will be shown below, $P_{\rm off}$ is always smaller than
$P_{\rm sky}$ in the solar observations due to the larger IF attenuation.
By analogy with Equation (1), our expression for the system temperature 
that can be compared
with values calculated by the ALMA on-line calibration system is

\begin{equation} 
T_{\rm sys}\ =\ {{P_{\rm sky}\,-\,P_{\rm zero}} \over {P_{\rm
hot}\,-\,P_{\rm amb}}} \ {{T_{\rm hot}\,-\,T_{\rm amb}} \over
{\eta_{\rm l}\,{\rm e}^{-\tau}}}
\end{equation}

\noindent where we correct for the atmosphere and forward efficiency
since those corrections are applied to the
$T_{\rm sys}$ values stored in ALMA datasets.

\begin{figure}[t]
\epsfxsize=150mm
\centerline{\epsffile{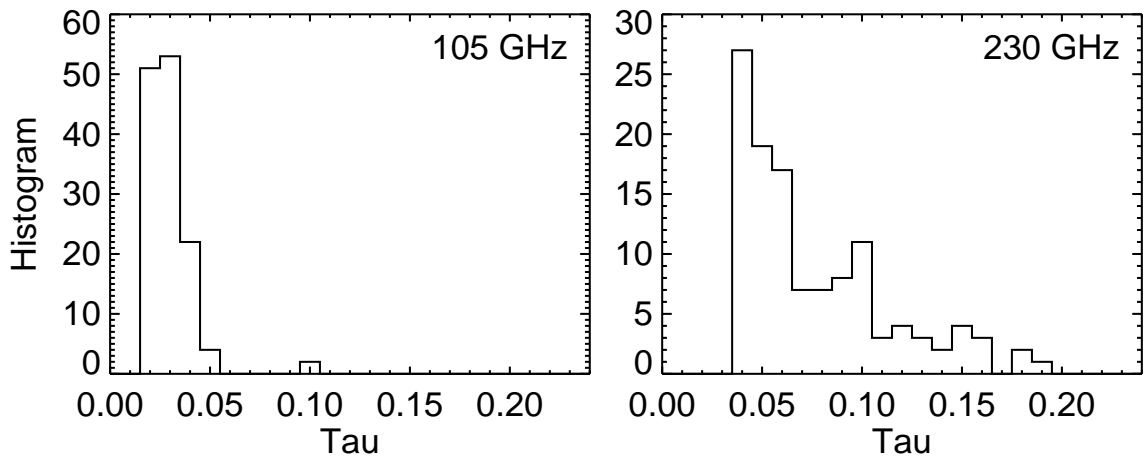}}
\caption{The distributions of daytime atmospheric opacity $\tau$ at 
Bands 3 (105 GHz, left panel) and 6 (230 GHz, right panel) extracted from
solar single-dish datasets in 2015 and 2016. 
}
\label{fig:tau}
\end{figure}

\subsection{Atmospheric Opacities}

As described earlier, ALMA uses measurements of the 183 GHz water-vapor
line together with an atmospheric model to infer the atmospheric opacity
and effective temperature required in the expressions above. The results of
fitting the 183 GHz water line by the WVR are supplied to the
atmospheric model, which is then used to calculate opacities at the
frequencies of each of the basebands being observed, 
so there are generally four values of $\tau$ provided with each dataset. 
At present, provision of these data for analysis 
requires connection of an antenna to an ALMA correlator, and
hence the WVR data from the individual PM antennas are not available
during solar observations since they are only using their total-power 
detectors and are not connected to a correlator.
In this case, the ALMA on-line system seeks a default water-vapor
measurement to use, usually from the nearest 12\,m antenna that is
connected to a correlator (as all the antennas in the interferometric
array are).  The PM antennas are often located on pads
close to the fixed array of 7\,m antennas, and frequently seem to be 
provided with identical sets of opacities. Figure \ref{fig:tau}
shows the typical range of daytime opacities measured at ALMA (at the
source elevation, not the zenith values) for one baseband
in each of Bands 3 and 6: at Band 3 $\tau$ is reliably below
0.05, whereas at Band 6 the smallest value is around 0.04 and the normal
spread reaches up to at least 0.15. In practice ALMA science
operations take $\tau$ into account when deciding what projects to run,
so at the higher values seen at Band 6 in Figure \ref{fig:tau}, Band 6
projects would probably not be run.

\subsection{ALMA System Calibration}

As described above, ALMA carries out a set of atmospheric-calibration
measurements at the beginning and end of each
fast-scanning single-dish dataset. These measurements are used by the
ALMA \textsf{TELCAL} software component \citep{BLP11} to
calculate the system temperature and atmospheric opacity in real time, 
and those measurements are passed on in the data to be used in the
production of the final calibrated map. 
The calibration measurements can be extracted from the data files using 
tools within \textsf{CASA}.

ALMA's system-temperature calculation employs the ``improved
dual-load'' method described by \citet{Luc12} and presented in 
the ALMA Technical
Handbook \citep[\eg section A.5.1 in][]{WaR17}, in which a combination of
the ambient and hot loads is used to approximate a load that matches
the temperature of the
atmosphere. In practice we find that the system temperatures
calculated by \textsf{TELCAL} and
reported by \textsf{CASA} are generally within $\pm$ 5\,\% of the single-load calculation,

\begin{equation} 
T_{\rm sys}\ =\ {{P_{\rm sky}\,-\,P_{\rm zero}} \over
{P_{\rm amb}\,-\,P_{\rm sky}}}\ {{T_{\rm amb}} \over
{\eta_{\rm l}\,{\rm e}^{-\tau}}}
\end{equation}

\noindent (discussed further below in reference to Table 4). The scaling
factor $g\,=\,T_{\rm sys}/P_{\rm ref}$ from Equation (1) needed to convert the target power to
a temperature measurement is then

\begin{equation}
g\ =\ {{T_{\rm sys}} \over {P_{\rm off}\,-\,P_{\rm zero}}}
\end{equation}

\noindent so that the single-load expression for calibration is

\begin{equation} 
T^*_{\rm src}\ =\ {{P_{\rm sky}\,-\,P_{\rm zero}} \over {P_{\rm
off}\,-\,P_{\rm zero}}}\
{{P_{\rm src}\,-\,P_{\rm off}} \over {P_{\rm amb}\,-\,P_{\rm sky}}}\
{T_{\rm amb} \over {\eta_{\rm l}\,{\rm e}^{-\tau}}}
\end{equation}

\subsection{Absolute Brightness Temperatures}

The result for $T^*_{\rm src}$ in (11) is not the final result that we
need for the measurement of solar atmospheric temperatures. A plethora of
temperature scales exists in the literature of millimeter-wave
astronomy \citep[\egb][]{Jew02,Man02},
and $T^*_{\rm src}$ is the source temperature on the telescope-dependent
``$T^*_A$'' scale  in which the antenna temperature is corrected for
atmospheric attenuation, radiative loss, and rearward scattering and
spillover \citep{KuU81,Man93}, but it does not account for
signal that enters the receiver from outside the subreflector, usually due
to the receiver pattern ``seeing'' beyond the edge of the subreflector. The
corresponding efficiency [$\eta_{\rm fss}$] is called the ``forward
scattering and spillover'' efficiency, 
and the resulting contribution [$1\,-\,\eta_{\rm fss}$] is therefore usually at
the temperature of the sky at the target elevation. 

To derive the actual brightness temperature of the Sun [$T_{\rm src}$] 
that would be seen by a perfect telescope above the
atmosphere, we need to divide $T^*_{\rm src}$ by $\eta_{\rm fss}$
\citep[\egb][]{Jew02}, 
so that

\begin{equation} 
T_{\rm src}\ =\ {{P_{\rm sky}\,-\,P_{\rm zero}} \over {P_{\rm
off}\,-\,P_{\rm zero}}}\ {{P_{\rm src}\,-\,P_{\rm off}} \over {P_{\rm
hot}\,-\,P_{\rm amb}}} 
\ {{T_{\rm hot}\,-\,T_{\rm amb}} \over {\eta_{\rm l}\,\eta_{\rm fss}\,{\rm e}^{-\tau}}}
\end{equation}

\noindent For an optically thick source such as the solar chromosphere, this
brightness temperature should be the actual temperature of the optically
thick layer producing the millimeter-wavelength
emission.\footnote{Technically this measurement is on the ``$T_R^*$''
temperature scale \citep{Jew02}, but since the Sun is so much larger than 
the beam the source-coupling
factor can be assumed to be unity and effectively this is the true
source brightness temperature.} In practice there will also be a
contribution to the measured brightness temperature from the 
optically-thin hot corona lying above the chromosphere, which could be
as much as a few hundred K in Band 3 from the densest parts of the
corona and scales with the inverse-square of frequency
\citep[\egb][]{Whi99a}.

\begin{table}[h]
\caption{Antenna efficiencies used for solar calibration. The final
column gives the correction factor 0.98/$\eta_{\rm l}$/$\eta_{\rm fss}$ to be
applied to solar single-dish images currently produced by standard
processing of ALMA data in \textsf{CASA}, which
assumes $\eta_{\rm l}$ = 0.98 and does not correct for $\eta_{\rm fss}$.}
\begin{tabular}{lccc}
\hline
        & Forward efficiency [$\eta_{\rm l}$] & Forward scattering and spillover [$\eta_{\rm fss}$] & \textsf{CASA} correction factor \\
\hline
Band 3 & 0.96 $\pm$ 0.02 & 0.91 $\pm$ 0.03 & 1.12 \\
Band 6 & 0.95 $\pm$ 0.03 & 0.89 $\pm$ 0.03 & 1.16 \\
\hline
\end{tabular}
\label{tab:eff}
\end{table}

\subsection{Antenna Efficiencies}

Unfortunately, neither $\eta_{\rm l}$ nor $\eta_{\rm fss}$
is well measured for the ALMA dishes. This does not affect calibration
of interferometric data since amplitudes are scaled to astronomical
calibrators; similarly, single-dish observations carried out with
the full sensitivity of normal receiver bias can also usually be
scaled to known bright calibrators such as planets. The \textsf{TELCAL} package that
determines system temperatures for ALMA observations uses a
fixed value $\eta_{\rm l}\,=\,0.98$ for all bands. This is likely to be
an overestimate of the true value, and it is unlikely that the same
value applies to all bands since different wavelengths can be expected to have
different scattering properties for a fixed antenna geometry.
\citet{TaS12} used sky-dip measurements, effectively just varying the
elevation and hence $\tau$ in Equation (2), to carry out
measurements of $\eta_{\rm l}$ that included the PM antennas. This is a
difficult measurement because the dependence of Equation (2) on $\eta_{\rm l}$ is weak
when $\tau$ is small, as it usually is for Bands 3 and 6, and any errors
in $T_{\rm rec}$ will affect the results. There is
considerable variation across the antennas measured. The results
are average values for the PM antennas of 0.94 at Band 3 and 0.92 at
Band 6, with systematic uncertainties of at least 5\,\%. 
These values are smaller than the nominal ALMA System Technical Requirement 
of $\eta_{\rm l}\,>\,0.95$.
\citet{TaS12} note that limiting the data to higher elevation
measurements, where all of the rear spillover comes from the ground as
assumed, tends to increase the resulting values.
Since the \textsf{TELCAL} value of 0.98 is believed to be too
large and the measurements of \citet{TaS12} may be slightly low, 
in the absence of better data we choose to minimize the uncertainty in
the value that we apply by using the average of these results and the
\textsf{TELCAL}
value of 0.98. This produces the values listed in Table \ref{tab:eff}: the
uncertainties associated with the choice of these values are unfortunate 
but we have no better measurements at present.

For $\eta_{\rm fss}$ there are two components to be considered. The ``forward
spillover'' results from the receiver being able to see the sky beyond
the edges of the subreflector on a Cassegrain telescope. Measurements of
forward spillover have been carried out during commissioning and from a
table of results provided by P. Yagoubov (private communication 2016), the MELCO antennas have an
average spillover value of 0.04 $\pm$ 0.01 at Band 3 and 0.06 $\pm$ 0.01
at Band 6. In addition,
blockage of the dish by the subreflector supports and scattering
from the central cone of the subreflector 
contribute an extra 5\,\% \citep[][M. Sugimoto and P. Yagoubov, private
communication, 2016]{SKI09}. Combining these two contributions gives the
values for $\eta_{\rm fss}$ reported in the middle column of Table
\ref{tab:eff}. Finally, since currently ALMA calibration assumes
$\eta_{\rm l}$=0.98 and the standard single-dish calibration using the 
$T_{\rm sys}$ values from \textsf{TELCAL} does not correct for $\eta_{\rm fss}$, the single-dish
images resulting from standard processing with ALMA calibration 
in \textsf{CASA} need to be corrected
to obtain the true solar brightness temperatures: the final column of
Table \ref{tab:eff} gives multiplication factors ($0.98/\eta_{\rm l}/\eta_{\rm fss}$) 
for this correction.\footnote{Note that in the standard \textsf{CASA} single-dish
calibration script this correction is applied as a ``gain'' in the
\textsf{gencal}
task. Since gains refer to values for each antenna that are then 
multiplied together in pairs to correct visibilities between two 
antennas, in the case of single-dish 
data the gain supplied is the inverse square-root of the
correction factor. Alternatively, the correction factor can be applied
directly after the imaging step with the \textsf{immath} task.}
Efforts continue at ALMA to provide better measurements of these
quantities, which may change the scaling of the data in the future.

An independent means of calibrating the Sun's brightness temperature that
avoids the need for detailed knowledge of antenna-beam efficiencies is to
use the Moon as a flux standard \citep[\egb][]{Lin73b,Lin73a}. The solid angle of
the Moon nearly matches that of the Sun and so measurements of the Moon
can be compared directly with those of the Sun. 
Use of the Moon for
calibration relies on the accuracy of lunar models for the time-,
frequency- and
position-varying response of the lunar surface to solar illumination. 
This undertaking
is significant and will be investigated in a future article.

\section{Fast-Scanning Solar Data from ALMA}

\subsection{Calibration Example}

\textsf{CASA} does not currently implement the calibration scheme
represented by Equation (16). In order to investigate the temperatures produced
by this method, we derived all of the data required to implement Equation (16) from 
the \textsf{CASA} data files using tools to extract table contents. An example of
the resulting calibration data, from a Band 6 dataset, is shown in Table
\ref{tab:cal}. The top five rows show the power measurements for each of
the five calibration points described earlier, for each baseband
and polarization. The values shown are averages of the middle section of
each calibration scan, representing about 2500 one-millisecond samples.
The standard deviation over this set of samples is 0.1\,\% or less of
the average values for all measurements except $P_{\rm zero}$, so the third digit
is significant. Note that the $P_{\rm zero}$ values are mostly small compared
to the other calibration data, as expected with the MD2 bias, and the
standard deviation of the data used to determine $P_{\rm zero}$ is at the
10$^{-4}$ level on the scale shown in Table \ref{tab:cal}. However,
even with these small values, for polarization Y at 248 GHz (baseband 4) 
$P_{\rm zero}$ is
about 12\,\% of $P_{\rm off}$, and since ($P_{\rm off}\,-\,P_{\rm zero}$) appears in
the denominator of (16), neglect of $P_{\rm zero}$ can result in about a
10\,\% difference in the temperature scale in this case, which would be
important for solar studies.  

\begin{table}[h]
\caption{Calibration measurements from a representative Band 6 dataset
using antenna PM04. The first five rows are median power
measurements for each of the calibration points described in
Section \ref{sec:calmeas}, in arbitrary units, for each baseband
(four frequencies, as labelled) and linear polarization (X, Y). The
$\tau$-measurements are as reported by the on-line system, while the
receiver temperature $T_{\rm rec}$ is calculated using Equation (9) and the sky
temperature contribution $T_{\rm sky}$ is calculated using Equation (10). The
``dual-load'' system temperature $T_{\rm sys}^{\rm dual}$ is given by
Equation (12), the
single (ambient) load system temperature $T_{\rm sys}^{\rm ambient}$ by
Equation (13),
while $T_{\rm sys}^{\rm Atmcal}$ reports the actual values used by ALMA for 
calibration, extracted from the data sets. In
order to illustrate the results of different calibration methods, the
final rows report the median brightness temperature in an
80\arcsec$\times$80\arcsec\ region centered at apparent disk center for
each of the three calibration approaches. T$_{\rm B}^{\rm dual}$ is the result of
a direct application of Equation (16), T$_{\rm B}^{\rm ambient}$ results from using the
single-load calibration Equation (15) with the correction for $\eta_{\rm fss}$, 
while T$_{\rm B}^{\rm Atmcal}$ results from
processing the data with \textsf{CASA} and then correcting for the chosen values
of $\eta_{\rm l}$ and $\eta_{\rm fss}$.
}
\begin{tabular}{lrrrrrrrr}
 & \\
\hline
 & \multicolumn{2}{c}{230 GHz} & \multicolumn{2}{c}{232 GHz} & \multicolumn{2}{c}{246 GHz} & \multicolumn{2}{c}{248 GHz} \\
 & X & Y & X & Y & X & Y & X & Y \\
\hline
Hot & 0.629 & 0.560 & 0.579 & 0.565 & 0.586 & 0.494 & 0.593 & 0.517 \\
Ambient & 0.590 & 0.534 & 0.545 & 0.539 & 0.553 & 0.472 & 0.560 & 0.493 \\
Sky & 0.456 & 0.447 & 0.426 & 0.453 & 0.437 & 0.402 & 0.446 & 0.417 \\
Off & 0.257 & 0.319 & 0.254 & 0.317 & 0.268 & 0.311 & 0.267 & 0.329 \\
Zero & 0.002 & -0.001 & 0.009 & -0.003 & 0.007 & -0.011 & 0.007 & -0.038 \\
\hline
$\tau$ & 0.142 & 0.142 & 0.156 & 0.156 & 0.156 & 0.156 & 0.182 & 0.182 \\
$T_{\rm rec}$ [K] & 729 & 1058 & 754 & 1092 & 790 & 1195 & 807 & 1176 \\
$T_{\rm sky}$ [K] & 57.8 & 71.2 & 60.9 & 72.1 & 61.6 & 74.4 & 65.3 & 80.2 \\
\hline
$T_{\rm sys}^{\rm dual}$ [K] & 955 & 1370 & 1003 & 1432 & 1048 & 1562 &
1101 & 1585 \\
$T_{\rm sys}^{\rm ambient}$ [K] & 1191 & 1814 & 1268 & 1904 & 1329 & 2097 & 1420 & 2189 \\
$T_{\rm sys}^{\rm Atmcal}$ [K] & 1019 & 1575 & 996 & 1478 & 1041 & 1621 & 1159 & 1826 \\
\hline
Disk $T_{\rm B}^{\rm dual}$ [K] & 5955 & 5979 & 5963 & 5990 & 5876 & 5907 & 5889 & 5903 \\
Disk $T_{\rm B}^{\rm ambient}$ [K] & 6124 & 6523 & 6128 & 6471 & 6056 & 6447 & 6012 & 6453 \\
Disk $T_{\rm B}^{\rm Atmcal}$ [K] & 6507 & 7107 & 5907 & 6454 & 5874 & 6548 & 6226 & 7820 \\
\hline
\end{tabular}
\label{tab:cal}
\end{table}

We carry out the calculations for each of the four observed frequencies 
and each of the two linear polarizations separately: we argue that 
all eight variants in a dataset should give similar results since 
solar emission is not linearly polarized and the relatively small frequency 
differences between the four basebands 
are not large enough to show up as significant temperature
differences in the final results. The fact that the X- and Y-maps
should give the same temperature at a given frequency is used by
\citet{SBH17} to determine the noise level in solar interferometry maps.
Table \ref{tab:cal} shows the typical
level of variation seen in the calibration, resulting in large
variations in both the receiver temperature [$T_{\rm rec}$] and the system
temperatures [$T_{\rm sys}$] across the eight samples. The system temperature
from Equation (12) derived using the two loads is often lower than both the
single-load system temperature from Equation (13) and the values
supplied to \textsf{CASA} by \textsf{TELCAL}
following the \citet{Luc12} approach. The two
latter values are generally similar but not identical to each other.
As expected, all values are around 1000 K rather than the values of
40\,--\,50 K typically found with the normal receiver bias.

\begin{figure}[t]
\epsfxsize=150mm
\centerline{\epsffile{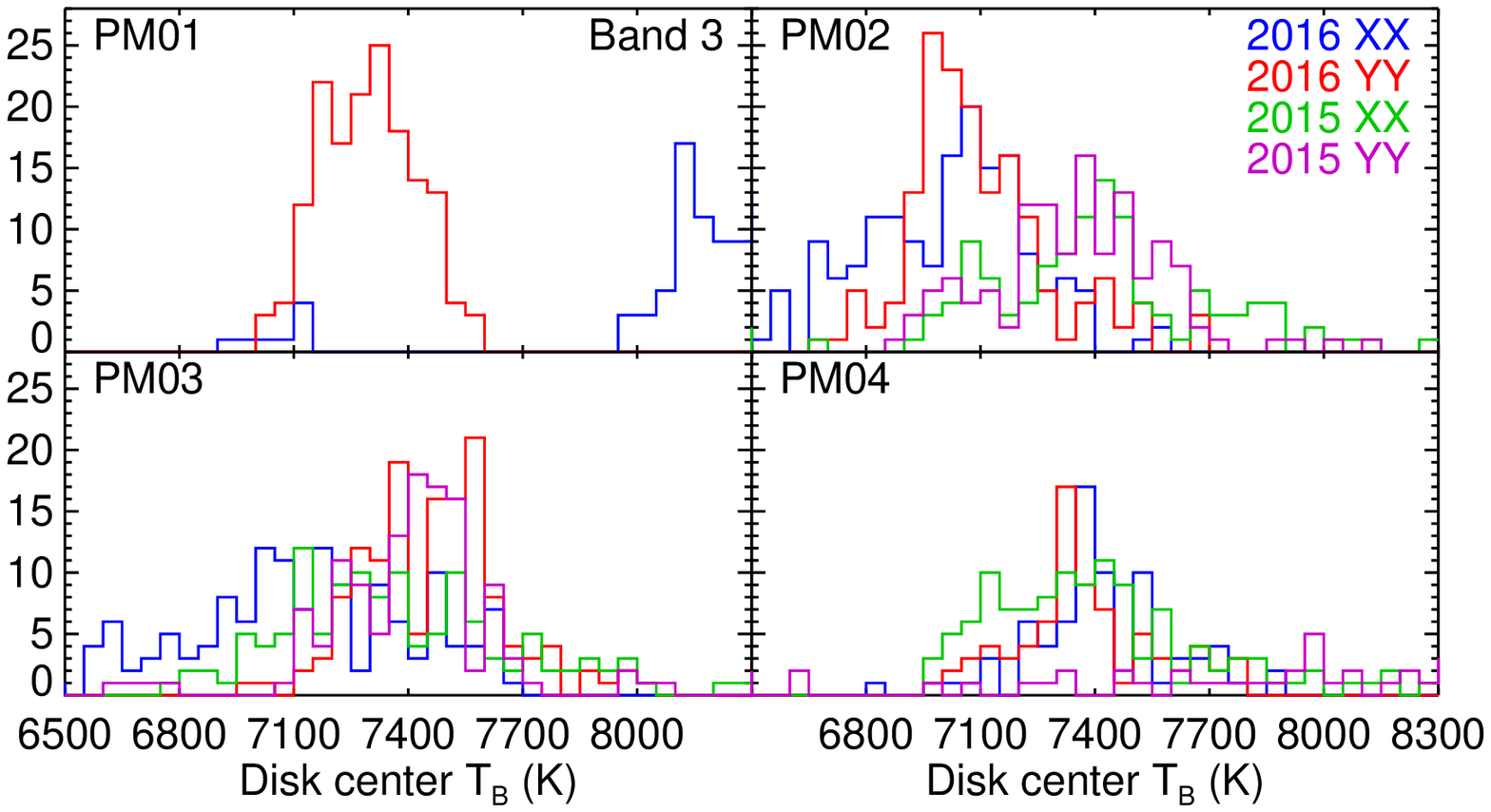}}
\caption{The distributions of the median brightness temperature obtained
in a 120\arcsec $\times$120\arcsec\ region at disk center (beam size
$\approx$ 60\arcsec) for Band 3
datasets from 2015 and 2016. The data are separated by antenna, year, and
polarization, as labelled. All four basebands (see Table
\ref{tab:freqs}) are included as separate values in the histograms. Note that there were
no measurements with PM01 in 2015, and as noted in the text, PM01 and
PM03 have poor calibration in polarization Y in 2016, while PM04 is
poor at Y in 2015.
}
\label{fig:tb3}
\end{figure}

\begin{figure}[hb]
\epsfxsize=150mm
\centerline{\epsffile{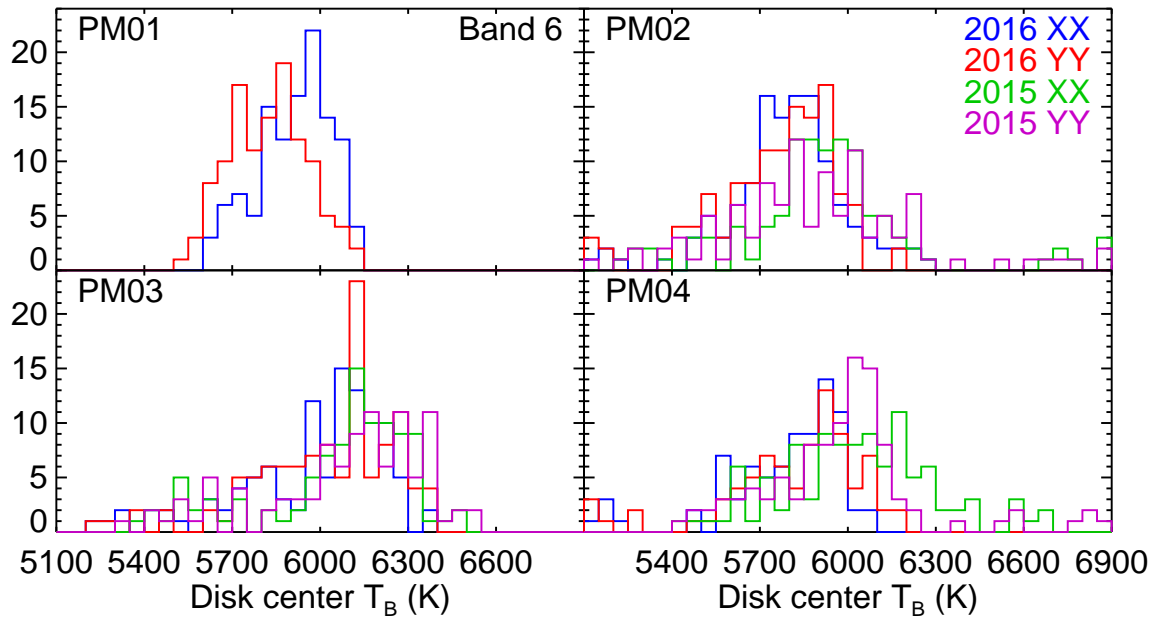}}
\caption{The distributions of the median brightness temperature obtained
in a 80\arcsec $\times$80\arcsec\ region at disk center (beam size
25\arcsec) for Band 6
datasets from 2015 and 2016. The data are separated by antenna, year, and
polarization, as labelled. All four basebands (see Table
\ref{tab:freqs}) are included as separate values. Note that there were
no measurements with PM01 in 2015.
}
\label{fig:tb6}
\end{figure}

Finally, the last three rows of Table \ref{tab:cal} compare the
brightness temperatures at the apparent center of the solar disk derived
using the three calibration schemes. The first set of values is derived
by application of Equation (16), the second by application of Equation (15) using the
single-load method with correction for efficiencies, 
while the last row results from processing the
data through \textsf{CASA} and applying the correction factor given in Table
\ref{tab:eff} to the resulting images. The dramatic result of this
comparison is that the eight values resulting from application of Equation (16) are
very close to each other despite the fact that the corresponding system
temperatures have a significant range: the disk center temperatures are
consistent to within better than 1\,\% in this case, even though the
system temperatures have a standard deviation of 20\,\% about their mean.
Much of this variation is due to a consistent difference between the two
polarizations: in all of the calibration methods the system temperatures of
all four sets of Y-polarization measurements are several hundred K larger
than those of the X-polarization measurements. This difference shows up
strongly in the solar temperatures calibrated with the single-load and
the version produced by the standard ALMA path through \textsf{CASA}: the Y-polarization maps
give much larger temperatures than the X-polarization maps.
Accordingly, the values resulting from standard ALMA calibration appear
to be less reliable than those obtained using Equation (16), and work is underway
to reconcile these differences.

\subsection{Solar Brightness Temperatures}

Although chosen at random, the above example of solar data is not
necessarily typical of an arbitrary solar dataset. In order to establish
the characteristics of solar fast-scanning data, a large number of
datasets from 2015 and 2016 have been processed, and it is frequently
found that the calibration using Equation (16) does not produce consistent
results. The datasets consist of the commissioning data from the 
December 2015 campaign and test datasets from the second half of 2016, when
solar activity was much lower than during the 2015 campaign.
It is found to be common that one polarization gives
systematically different results from the other, particularly for a
given antenna at a given band. For example, in 2015, PM04 had trouble measuring
$P_{\rm zero}$ in polarization Y at Band 3, so those brightness temperatures vary
wildly, whereas the X brightness temperatures were much more consistent. 
In 2016, both PM01 and PM03 produced consistently excessive brightness 
temperatures in polarization X
at Band 3 while the corresponding Y brightness temperatures were more
normal.

In order to assess the robustness of solar brightness temperatures
obtained using Equation (16) with the ALMA calibration measurements, Figures
\ref{fig:tb3} and \ref{fig:tb6} show the distributions of disk-center
brightness temperatures for a number of datasets at Bands 3 and 6,
respectively, separating the data by PM antenna, year, and polarization.
As far as we know there are no active regions in the low-latitude
region around disk center in any of the datasets, so we believe that they
should all represent quiet-Sun chromosphere with presumably similar
median brightness temperatures, averaged over areas large compared to
the typical spatial scales of the network structure likely to be
present. We therefore argue that the true brightness temperature at disk
center should be very similar in all datasets at a given band, and the
spread seen in Figures \ref{fig:tb3} and \ref{fig:tb6} represents the
level of uncertainty introduced by the calibration scheme. We note that
100 K error is less than 2\,\% of the absolute temperatures in the range
6000\,--\,7000 K, but since the goal of the single-dish data is to provide
the absolute temperature scale for ALMA solar data, we wish to provide
the best possible guidance for solar brightness temperatures.

\begin{figure}[th]
\epsfxsize=150mm
\centerline{\epsffile{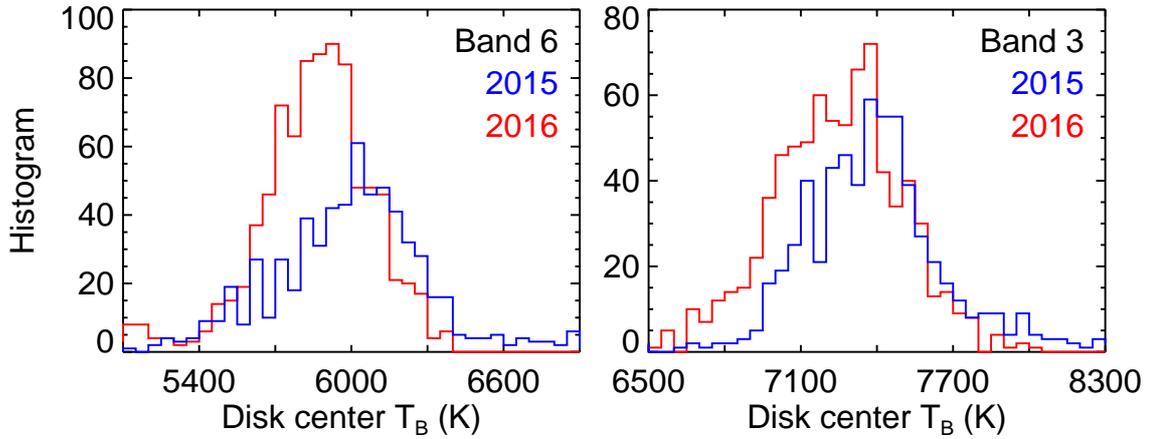}}
\caption{The distributions of the disk center brightness temperatures at
Bands 6 (left panel) and 3 (right panel) for all four PM antennas, merged,
and excluding datasets known to be bad. The distributions for 2015
(blue) and 2016 (red) are plotted separately.
}
\label{fig:tb}
\end{figure}

\begin{figure}[b]
\epsfxsize=150mm
\centerline{\epsffile{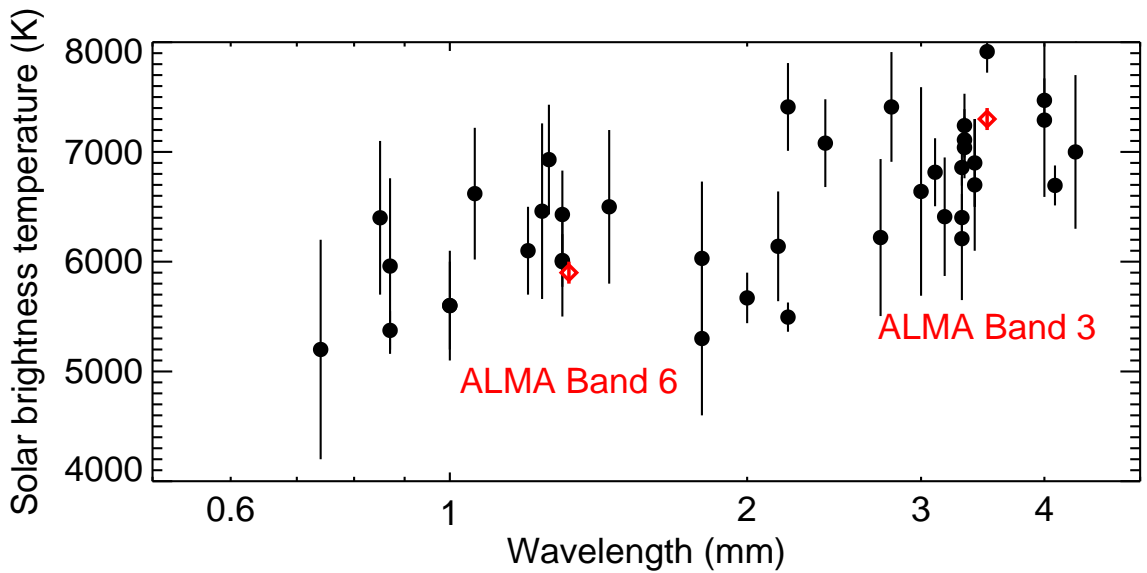}}
\caption{The recommended ALMA disk-center brightness temperatures (red
diamonds with $\pm$100 K error bars) at
Bands 6 (1.3 mm wavelength) and 3 (3.5 mm), together with previous
measurements (black circles with measurement uncertainties) from the 
compilation by \citet{LSC15}.
}
\label{fig:meas}
\end{figure}

To that end, in Figure \ref{fig:tb} we plot the distributions of
brightness temperatures for Bands 3 and 6 in 2015 and 2016, separately,
excluding data groups identified as bad from Figures \ref{fig:tb3} and
\ref{fig:tb6} as described above. We have then fitted Gaussians to the
distributions with the results shown in Table \ref{tab:tb}. In both
bands, the values for 2015 are on average about 100 K brighter than in
2016. We are unsure whether this represents a true difference
resulting from changes in the chromosphere associated with the lower
solar-activity levels in 2016, or whether this is due to the overall
uncertainty in the calibration. At higher activity levels, a larger 
relative contribution of hot network features compared to cooler cell 
interiors could certainly explain higher quiet-Sun levels at the
single-dish spatial resolution. In any case, the central values are
sufficiently similar for us to recommend that the quiet-Sun disk-center
brightness temperature of the Sun be set to 7300 K at 100 GHz and 5900 K at 230
GHz. The formal standard error in the fitted central values associated
with the statistical spread of results is much smaller than the uncertainty 
of 100 K ($\approx$1.5\,\%) that we quote: we use that value to account for 
the difference between the 2015 and 2016 results. 
The uncertainty of 100 K does not account for the systematic uncertainty 
associated with the choice of values for $\eta_{\rm l}$ and $\eta_{\rm fss}$ 
(Table \ref{tab:eff}), which is much larger (5\,\%) than this value. 
Given the large spreads shown in Figure
\ref{fig:tb}, and the uncertainty in the calibration provided by current \textsf{CASA}
processing of the single-dish solar data, it is our recommendation that
single-dish solar maps should be scaled to these values.

\begin{table}[h]
\caption{Central brightness temperatures and the spread in values for 
Bands 3 and 6 resulting
from fitting Gaussians to the distributions shown in Figure \ref{fig:tb}.
For the first two lines the uncertainty is the Gaussian width of the fit.
The third line lists the recommended values for the two bands.
}
\begin{tabular}{l|cc}
\hline
        & Band 3 & Band 6 \\
\hline
2015 & 7390 $\pm$ 220 K & 6040 $\pm$ 250 K \\
2016 & 7280 $\pm$ 250 K & 5900 $\pm$ 190 K \\
Recommended & 7300 $\pm$ 100 K & 5900 $\pm$ 100 K \\
\hline
\end{tabular}
\label{tab:tb}
\end{table}

We can compare these numbers with previous measurements. \citet{LSC15}
provide a compilation of historical brightness temperature measurements of 
the Sun at millimeter wavelengths. We replot those data here in Figure
\ref{fig:meas}, together with the recommended values from Table
\ref{tab:tb}. Note that the values plotted are not all directly
comparable, since different measurements had different spatial
resolution and therefore were averaging over different areas of the
solar surface. As discussed in the next section, other effects such as
limb brightening will also play a role in the measured values: our
recommended values apply specifically to quiet-Sun regions at apparent
disk center, at the resolution of the ALMA single-dish data. 
The ALMA values are consistent with previous data, but
with much smaller uncertainty due to the large number of values involved
in our assessment. 

\begin{figure}[t]
\epsfxsize=160mm
\centerline{\epsffile{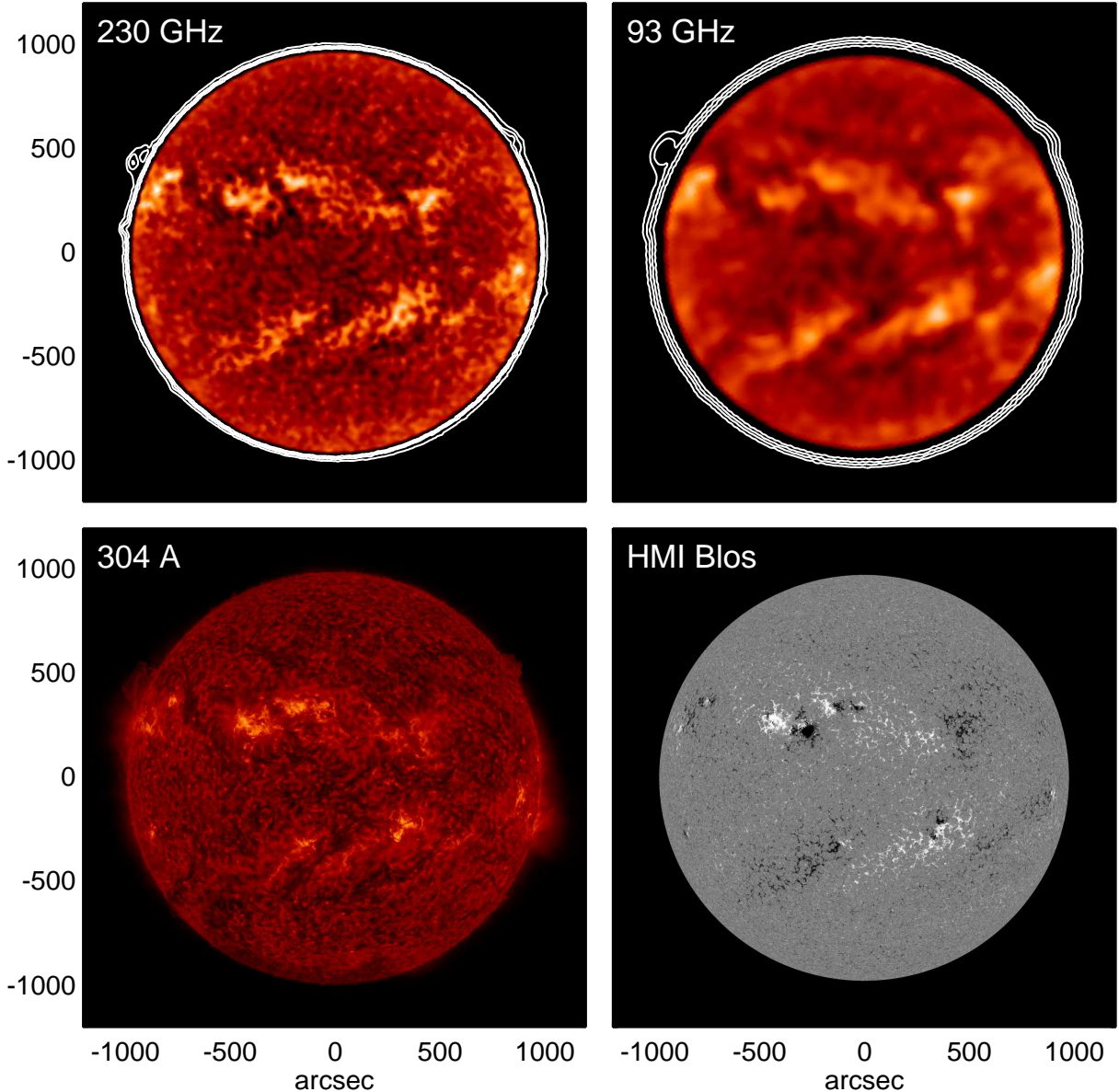}}
\caption{ALMA fast-scanning observations of the Sun on 17 December 2015
at Band 6 (230 GHz at 14:33 UT), upper-left panel) and Band 3 
(93 GHz at 18:15 UT, upper-right panel). In order to
emphasize structure on the disk, the 230 GHz image color display ranges
from 5300 to 7400 K, while the 93 GHz color display ranges from 6700 to
8800 K. Low-level contours are plotted at 300, 600, 1200, and 2400 K in
order to show features above the limb. For comparison, we show a {\it Solar
Dynamics Observatory} (SDO)
{\it Atmospheric Imaging Assembly} image of the Sun at 304 \AA\ 
(lower-left panel, dominated by the He{\sc ii} line from the upper chromosphere, 
at 14:44 UT to match 230 GHz), and a line-of-sight magnetogram from the
{\it Helioseismic and Magnetic Imager} (HMI) on board SDO at 18:30 UT (lower-right
panel).
}
\label{fig:ims15}
\end{figure}

\citet{LSC15,LWS17} use state-of-the-art radiative hydrodynamic simulations
of the solar chromosphere \citep[from the \textsf{Bifrost} code;][]{CHG16} 
to predict brightness temperatures at millimeter wavelengths. The
spatial resolution of the simulations is subarcsecond, and so it is not
straightforward to compare their results with the much poorer spatial
resolution of the single-dish data. They find average brightness
temperatures of 5000 K at 230 GHz and 6200 K at 94 GHz, corresponding
to effective heights above the photosphere of about 1150 km at 230 GHz
and 1600 km at 94 GHz \citep{LWS17}. Earlier simulations of inter-network quiet-Sun
regions by \citet{WLS07} found even lower average brightness
temperatures. These temperatures are substantially
smaller than the values we find here, but detailed comparison
requires the use of ALMA interferometer data with a much better match to
the spatial resolution of the simulations. If it turns out that 
the efficiencies quoted in
Table \ref{tab:eff} are underestimates of the true values, then the
brightness temperatures that we determine with the ALMA observations will be
reduced.

\begin{figure}[t]
\epsfxsize=140mm
\centerline{\epsffile{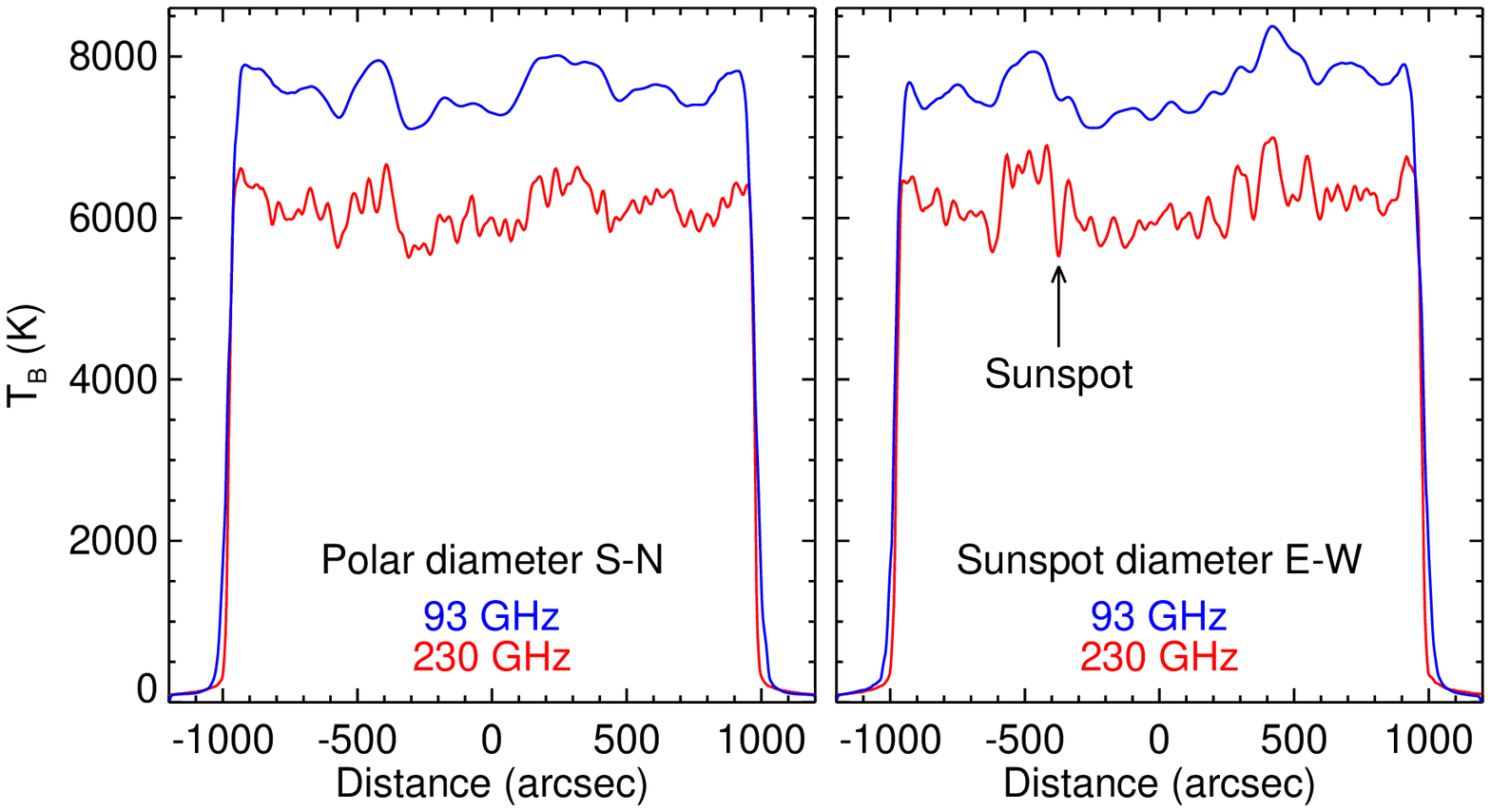}}
\caption{Profiles across the solar disk at 93 (blue) and 230 GHz (red) 
at two angles: from the South
Pole to the North Pole at apparent disk center, and on a diameter
through the large sunspot (azimuth 63.8\deg\ East of North), from East 
to West. The 93 GHz image has been rotated back to the time of the 
230 GHz image for the comparison. The location of the sunspot in the
right panel is marked by an arrow.
}
\label{fig:cuts}
\end{figure}

\subsection{Calibration Error Budget}

The sources of measurement error in the data required for the application of 
Equation (16) are as follows:

\begin{itemize}

\item The calibration power measurements [$P_{\rm hot}$, $P_{\rm amb}$,
$P_{\rm sky}$,
and $P_{\rm off}$] have standard deviations about the mean of order
0.1\,\%.
$P_{\rm zero}$ has a larger relative uncertainty, being limited by
16-bit quantization in the measurement, but it always appears as a
number to be subtracted from one of the other power measurements, and
its uncertainty of order 10$^{-4}$ is at the same absolute level as the
other measurements and therefore it too contributes less than 0.1\,\% to
the overall uncertainty.

\item The load temperatures $T_{\rm hot}$ and $T_{\rm amb}$ are measured and
their values are present in the datasets. The ALMA
technical requirement is an accuracy of 0.3 K for the ambient load and 1.0
K for the hot load \citep{YMW11}. Inspection of a large number of
datasets during the commissioning campaigns showed variability within a
1 K range, consistent with the above requirement and 
corresponding to an uncertainty of order 0.3\,\%. The
load temperatures can vary within about 10 K of their nominal values,
but the actual measured temperatures are used by \textsf{TELCAL}.

\item The atmospheric opacity [$\tau$] is determined by the WVR
measurements of the 183 GHz water line in
combination with an atmospheric model as
described earlier. We do not have a good estimate of the uncertainty in
its determination by the ALMA system. Previous studies have
compared the opacity measured by a tipper system, usually at 225 GHz,
with measurements of PWV and found a spread of order 0.01-0.02 in
opacity for a given value of PWV \citep[\eg ][]{CaB95,DOB99}.
In the calibration expressions $\tau$ appears as the factor
$\exp(-\tau)\,\approx\,1\,-\,\tau$, and for Bands 3 and 6 typically
$\tau\,<\,0.1$ and errors are unlikely to be larger than, say, 0.02. This
could give a several percent error in the resulting calibration.

\item The uncertainty in the measurements of $P_{\rm src}$ are not well
known. We can estimate an upper limit by investigating the variability
of the power as the double-circle pattern passes through disk center on
each pass: we find again that the variability, much of which is likely
to be real solar variability due to oscillation power in the 
chromosphere \citep[\egb][]{LSW06}, is typically less
than 2\,\% for Bands 3 and 6, and the different basebands and
polarizations agree to within about 0.1\,\%. This
variability is on the timescale of pattern passes through disk center,
which is of the order of a second or less. Fluctuations in atmospheric
opacity will also affect $P_{\rm src}$, but they generally have timescales of
minutes rather than the millisecond sampling time. Thus we do not expect that
the uncertainty in the $P_{\rm src}$ measurements is any worse than in the
calibration measurements.

\end{itemize}

\begin{figure}[th]
\epsfxsize=120mm
\centerline{\epsffile{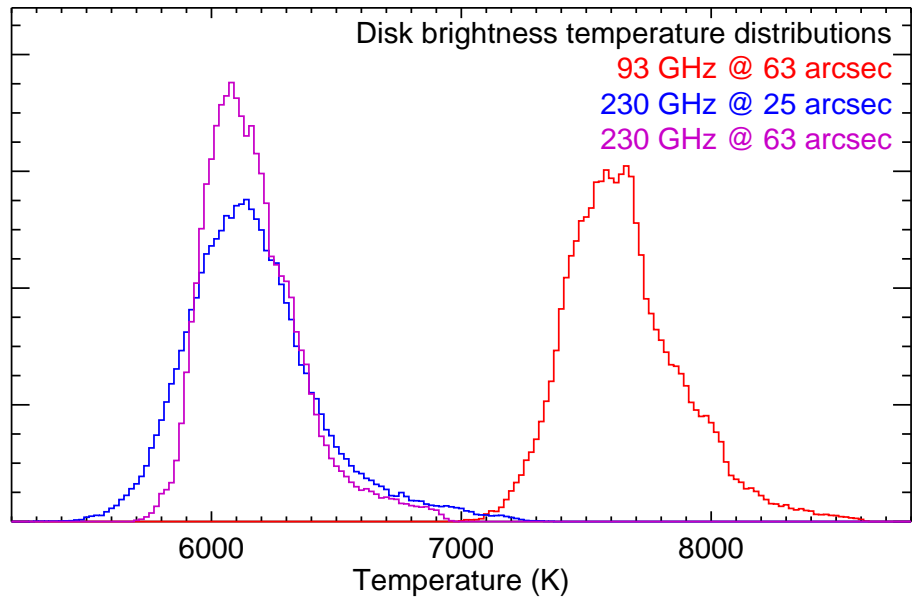}}
\caption{Histograms (arbitrary scale) of temperature distributions on the disk for the
images in Figure \ref{fig:ims15}. The plot includes positions out to 90\,\%
of the solar radius. In order to show the likely effects of spatial
resolution at 93 GHz, the distribution of 230 GHz temperatures when the
image is convolved to the 93 GHz resolution is shown (purple histogram), 
in addition to the distributions at the native resolution of the 93 GHz 
(63\arcsec, red histogram) and 230 GHz (25\arcsec, blue histogram) data.
}
\label{fig:histo}
\end{figure}

The combined set of uncertainties described above, in
particular from the calibration loads and the measurement of $\tau$,
yield an uncertainty of order 2-3\,\%.  The
5\,\% uncertainty in the product $\eta_{\rm l}\,\eta_{\rm fss}$ is an
uncertainty in the overall temperature scale: this product could vary from 
one PM antenna to another and explain differences in the centroids of the
individual antenna distributions seen in Figures \ref{fig:tb3} and
\ref{fig:tb6}, but it does not contribute to the spread
of results for each individual antenna. 
The nominal requirement for calibration of ALMA interferometer 
visibilities is 5\,\% (P. Yagoubov, private communication).
An overall uncertainty in the calibrated temperatures at the few
percent level is desirable for quantitative solar work in
order to study the energetics of the solar atmosphere, 
particularly in comparing results at different frequency bands
\citep[\eg ][]{WBB16}.

\subsection{Solar Fast-Scanning Images}

Representative examples of ALMA fast-scanning images of the Sun from 
17 December 2015 are shown in Figure \ref{fig:ims15}.
Corresponding examples of ALMA solar interferometer data are
presented by \citet{SBH17} in a companion paper. The single-dish
images in Figure \ref{fig:ims15} are
generated using \textsf{CASA} tasks implementing calibration and gridding as
described earlier.\footnote{Solar single-dish data should be processed
with \textsf{CASA} version 4.7 or later versions.} The brightness
temperature display range is chosen to show the full range of brightness
temperatures on the disk at Band 6 (left) and Band 3 (right), and
lower-level contours are plotted to show off-limb features. For
comparison, we show a 304 \AA\ image and a line-of-sight magnetogram
from instruments on board the {\it Solar Dynamics Observatory} (SDO). The 304 \AA\
image is dominated by the He{\sc ii} line and represents the upper
chromosphere. As expected, the bright regions in the ALMA images match
the bright regions at 304 \AA\ and the areas of strong magnetic field
where additional atmospheric heating would be expected. However, there
are obvious differences in the relative brightness of different
features: \eg the brightest feature in the south-east quadrant in the
ALMA images is not the brightest feature in the same area in the 304
\AA\ image. Further, there are significant differences in the relative
brightness of individual features at 230 and 93 GHz, \eg at the same
resolution, the active region plage East of the large sunspot is
relatively brighter at 230 GHz than at 93 GHz.
Prominences visible above the limb in the ALMA images match
off-limb emission in the 304 \AA\ image, although again with
differences in relative brightness.  A large sunspot is present in the 
north-east quadrant,
coincident with the intense negative polarity in the magnetogram. It is
clearly visible as a depression surrounded by a bright rim in the 230
GHz image, but it is less obvious at the 60\arcsec\ resolution of the 93
GHz image.  Large-scale cool features are present over filament channels
at both ALMA frequencies.

\begin{figure}[t]
\epsfxsize=160mm
\centerline{\epsffile{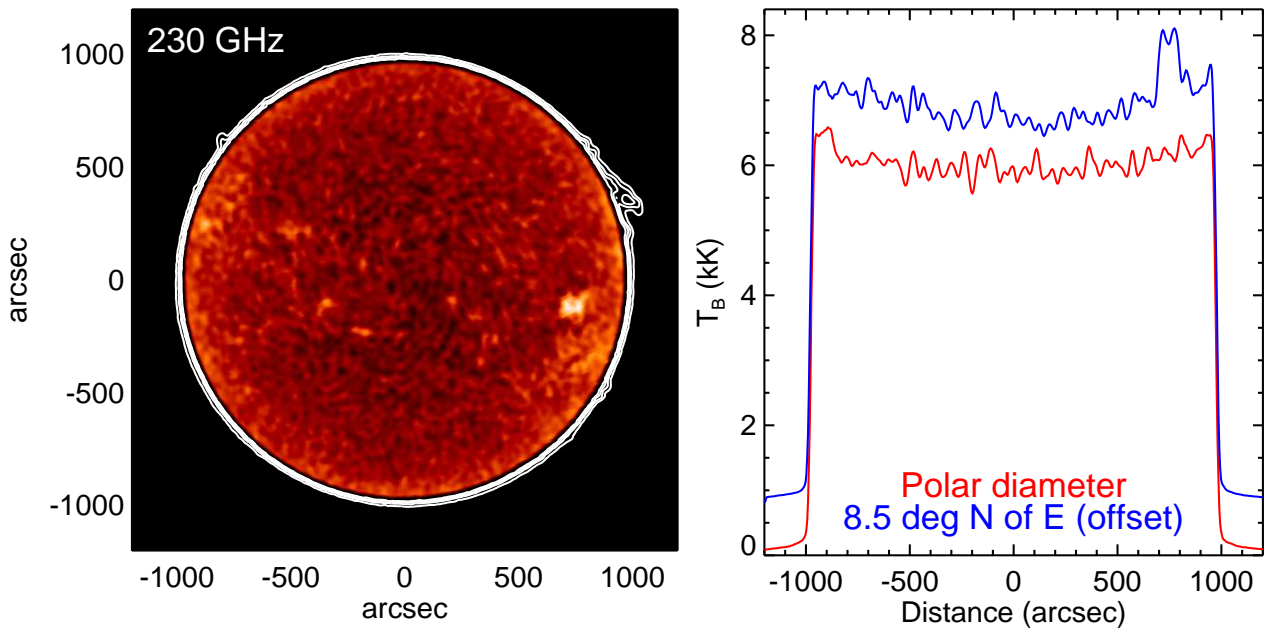}}
\caption{A 230 GHz image of the Sun on 7 December 2016, at
a much lower activity state than in Figure \ref{fig:ims15}. The display
range and the contours in the left panel are identical to those of the 
230 GHz image in Figure \ref{fig:ims15}. The right panel shows disk
profiles through the Poles and on a diameter through the active region in
the southwest quadrant, but with the blue
curve offset by 800 K in order to show structure in both.
}
\label{fig:ims16}
\end{figure}

To provide a more quantitative picture of the level of variability
across the disk, Figure \ref{fig:cuts} shows profiles through disk
center from both ALMA images along two angles: a polar cut from South to
North (left panel), and a diameter at 64\deg\ East of North that passes 
through the large sunspot (right panel). A direct comparison between the
93 and 230 GHz profiles is not possible due to the differing
resolutions at the two frequencies: the larger beam at 93 GHz is
averaging over about six resolution elements at 230 GHz. The polar cut is
dominated by quiet-Sun fluctuations amounting to a few hundred K on
small spatial scales. The sunspot diameter crosses a number of bright
regions, and in particular the active region encompassing the sunspot,
between about $-700$\arcsec\ and $-300$\arcsec, which is around 1000 K
brighter than the adjacent regions at 230 GHz. The dip in emission over
the sunspot umbra is clearly visible in the 230 GHz profile, being less
obvious at the resolution of the 93 GHz data: as discussed by
\citet{LSW14}, atmospheric models do not predict such a dip in the
brightness of the umbra at millimeter wavelengths.

Figure \ref{fig:histo} shows histograms of the temperature distributions
of the 93 and 230 GHz images in Figure \ref{fig:ims15} on the disk out to
90\,\% of the solar diameter. It can be seen that the 230 GHz temperatures
range from 5400 K up to 7300 K, while the 93 GHz temperatures range from
7000 to 8600 K. The larger temperature range at 230 GHz (almost 2000 K)
is likely a function of spatial resolution: when convolved to the
63\arcsec\ resolution of the 93 GHz map (purple histogram in Figure
\ref{fig:histo}), the range of
temperatures reduces to 5700\,--\,7000 K. The ``recommended'' temperatures
for quiet-Sun at disk center are on the low side of the histograms in
both cases, as expected with the large area of more active Sun present
on the disk.

By contrast, Figure \ref{fig:ims16} shows a 230 GHz image from a year
later at a much lower level of solar activity. Only one active region is
visible on the disk, and again it is almost 1000 K brighter than the
surrounding quiet Sun. In this case the profiles across the disk show much
lower levels of variability. The quiet-Sun network structure is very
prominent in this image, and the contrast between network and cell
interiors dominates the variability on small spatial scales. 
Prominences above the limb are again readily visible.
In addition, there is clear evidence for
limb brightening in this image, which might also be present in the 2015
data but could be masked there by the higher overall level of activity. 

Unfortunately, these images are not sufficient to study the limb
brightening and compare with chromospheric models because they are
convolved with the point-spread response of the single-dish telescope,
which has broad wings: the effects of these wings are evident in the
relatively smooth fall-off of emission above the limb (beyond
1000\arcsec\ from disk center) in Figures \ref{fig:cuts} and
\ref{fig:ims16}. The wings of the point-spread function (PSF) will play a role
in the apparent amount of limb brightening, and deconvolution of the PSF
from the images is required for quantitative analysis. This effect will be
addressed in a future study.

\section{Conclusions}

We have described solar single-dish imaging using the fast-scanning 
technique implemented on 12\,m ALMA dishes. The results of commissioning
are summarized, leading to the availability of single-dish maps to
supplement ALMA interferometer observations of the Sun. The
single-dish data provide the absolute temperature scale missing from
the interferometer data, and can be used to fill in the large-scale
structure when deconvolving mosaic images \citep[\egb][]{SBH17}. 
The observations are carried out 
in a ``detuned'' receiver mode in which the mixers are biassed to a level that
provides a lower gain with a more linear response over the full
temperature range appropriate to the Sun. A calibration path using all 
five calibration
power measurements, together with atmospheric properties from the WVR
and the ``forward'' [$\eta_{\rm l}$] and ``forward scattering and spillover''
[$\eta_{\rm fss}$] antenna efficiencies, is shown to be successful in 
determining the solar temperature. This calibration path is not yet 
implemented in \textsf{CASA}. The uncertainty in the calibration of
individual datasets is larger than is desirable for solar applications,
and we recommend
that the standard ALMA single-dish image products be scaled to match our best
determination of quiet-Sun temperatures at disk center
(specifically, 5900 K averaged over an 80\arcsec-square 
region centered on apparent disk center at Band 6, and 7300 K averaged over a
120\arcsec-square region centered on apparent disk center at Band 3, with 
uncertainties of order 100 K) while work 
continues to understand the source of 
this variation. There is uncertainty in the exact values of the two antenna
efficiencies required for calibration, which affects the scaling of the
data and the recommended values above, and better measurements of these
properties may be available in the future.

Analysis of sample images shows that the temperature on the disk can
vary over about a 2000 K range at the 25\arcsec\ resolution of the Band
6 data: this implies that an even larger range of variation should be
seen in the sub-arcsecond interferometer images. Active
regions, plage, sunspot umbrae, prominences, and filament channels are all 
strikingly present in the images, demonstrating the range of features
and properties of the solar atmosphere that ALMA will study.

%

%

%

%
\begin{acks}

The ALMA solar commissioning effort was supported by ALMA Development
grants from NRAO (for the North American contribution), ESO (for the
European contribution) and NAOJ (for the East Asia contribution). 
The help and cooperation of the ALMA Extension and Optimization of
Capabilities (EOC) team as well as the engineers, telescope
operators, astronomers-on-duty and staff at the ALMA Operations
Support Facility was crucial for the success of solar commissioning
campaigns in 2014 and 2015. We are grateful to the ALMA project for
making solar observing with ALMA possible.
R. Braj\v{s}a acknowledges partial support of this work by Croatian Science
Foundation under the project 6212 ”Solar and Stellar Variability” and by the
European Commission FP7 project SOLARNET (312495, 2013 - 2017), which is an
Integrated Infrastructure Initiative (I3) supported by the FP7 Capacities
Programme. G. Fleishmann acknowledges support from NSF grants AGS-1250374 and AGS-1262772.
Travel by Y. Yan to ALMA for the 2015 commissioning campaign was partially 
supported by NSFC grant 11433006.

\end{acks}

\section*{Disclosure of Potential Conflicts of Interest}

The authors declare that they have no conflicts of interest affecting
this article.


%
%
%

\end{article} 
\end{document}